\documentclass[a4paper,showpacs,nofootinbib]{revtex4-2}
\usepackage[utf8]{inputenc}
\usepackage{graphicx,bm,bbm,color,amssymb,amsmath}
\usepackage{subcaption}
\usepackage{slashed,verbatim}
\usepackage{soul}
\usepackage[usenames,dvipsnames]{xcolor}
\sethlcolor{red}
\usepackage[colorlinks=true,linktocpage=true,linkcolor=blue,citecolor=blue]{hyperref}
\usepackage{ulem}
\usepackage{cancel}  

\newcommand{\sF}{\scriptscriptstyle{f}}

\newcommand{\sperp}{\scriptscriptstyle{\perp}}
\newcommand{\shp}{\shortparallel}
\newcommand{\wt}[1]{\widetilde{#1}}
\def\nn {\nonumber}

\begin{document}
\title{Dilepton Spectra and Even Flow Harmonics in a Magnetized QGP: An Ideal Hydrodynamic Study}
\author{Ankit Kumar Panda$^{1,a}$, Aritra Das$^{2,b}$, Ashutosh Dash$^{3,c}$, Aritra Bandyopadhyay$^{4,d}$, Chowdhury Aminul Islam$^{5,e}$}

\affiliation{$^1$School of Physical Sciences, National Institute of Science Education and Research, An OCC of Homi Bhabha National Institute, Jatni-752050, India}
\affiliation{$^2$Department of Physics and Astronomy, Iowa State University, Ames, Iowa, 50011, USA}
\affiliation{$^3$Institute for Theoretical Physics, Goethe University, Max-von-Laue-Str.1, D-60438 Frankfurt am Main, Germany}
\affiliation{$^4$Department of Physics, West University of Timişoara, Bd. Vasile Pârvan 4, Timişoara 300223, Romania}
\affiliation{$^5$Center for Astrophysics and Cosmology, University of Nova Gorica, Vipavska 13, SI-5000
Nova Gorica, Slovenia}

\email{$^a$ankitkumarpanda932@gmail.com}
\email{$^b$aritrad@iastate.edu}
\email{$^c$dash@itp.uni-frankfurt.de}
\email{$^d$aritra.bandyopadhyay@e-uvt.ro}
\email{$^e$chowdhury@guest.ung.si}

\begin{abstract}
We present the first comprehensive study of dilepton production from a hot, magnetized quark-gluon plasma in heavy-ion collisions (HIC), incorporating realistic, time-dependent, and spatially inhomogeneous magnetic field profiles within an analytically solvable Gubser flow background. This framework provides a significant improvement over previous static calculations with homogeneous fields and moves toward the long-term goal of full $3+1$D magnetohydrodynamic simulations. We explore the effects of impact parameter, electrical conductivity, and invariant mass on the dilepton spectra and anisotropic even flow coefficients. It is found that transverse momentum spectra increase with impact parameter, dominated by annihilation processes, while decay contributions remain sub-leading. Strikingly, the elliptic flow $v_2$ from decay channels is nonzero even in nearly central collisions, exhibiting a characteristic shape\textemdash positive at low $p_T$ and negative at high $p_T$\textemdash that is largely independent of impact parameter and conductivity. In contrast, $v_2$ from annihilation processes is smaller in magnitude but dominates the total flow in magnitude due to its larger yield. Higher harmonics, such as $v_4$, are an order of magnitude smaller as compared to $v_2$ along with distinctive zero-crossing patterns. Conductivity enhances both spectra and flow but leaves no unambiguous signature for its extraction. Varying the invariant mass reveals the strongest enhancements at low mass, with harmonic coefficients suppressed at higher masses. Overall, our results suggest that central and semi-central collisions can carry imprints of the background magnetic field, and that characteristic correlations in even flow harmonics may provide a robust probe of electromagnetic effects in HICs.
\end{abstract}

\maketitle
\newpage

\section{Introduction} 
\label{sec:intro}
Electromagnetic emissions such as photons and dileptons are considered among the most reliable probes of quark-gluon plasma (QGP) created in relativistic heavy-ion collisions (HIC)~\cite{Wong:1994cy,Yagi:2005yb}. They, being electromagnetically interacting particles, escape the medium (which is essentially a strongly interacting system) with minimal final-state interactions, thereby carrying less contaminated and hence direct information about the thermodynamic and dynamical properties of the QGP. Along with this, in recent years, there has been a growing interest in understanding the impact of external magnetic fields on their production and evolution, particularly in the context of ultra-relativistic HICs, where intense magnetic fields are generated~\cite{Skokov:2009qp}. The presence of strong magnetic fields, typically of the order of $eB\sim m_\pi^2;~(m_\pi\approx 140$ MeV) or higher are produced, at the center of collision, in the early stages of HICs~\cite{Gursoy:2014aka,Dash:2023kvr,Panda:2024ccj}. This significantly modifies the in-medium spectral properties of quarks and antiquarks, affecting the production rate of photons or dileptons (virtual photons)~\cite{Kharzeev:2013jha,Miransky:2015ava,Mustafa:2025uad} and also its transport properties~\cite{Ghosh:2018cxb,Denicol:2018rbw,Denicol:2019iyh,Dash:2020vxk,Panda:2020zhr,Panda:2021pvq}.

This modification arises due to Landau quantization and anisotropic screening effects, which influence the quark propagators and, consequently, the electromagnetic current-current correlation function. Several efforts have been made to estimate the modifications for both photon and dilepton production. It has also been studied that these strong fields may lead to particle production via the Schwinger effect~\cite{Tanji:2008ku,Gould:2019myj,Gould:2021bre}. Incorporating these magnetic fields into the theoretical frameworks in order to estimate electromagnetic emissions was challenging. This led to estimations of the rate in approximate forms—either in terms of the strength of the magnetic field (for example, the lowest Landau level (LLL) approximation in the regime of strong magnetic field) or by considering specific values of external parameters such as momentum. Such approximate results for the dilepton rate using formal field theoretical approach can be found in~\cite{Sadooghi:2016jyf,Bandyopadhyay:2016fyd,Bandyopadhyay:2017raf,Ghosh:2018xhh,Das:2019nzv,Hattori:2020htm}, where both the Ritus eigenfunction method and the Schwinger proper time method were used. 

A general expression for the dilepton production rate in the presence of magnetic fields was first derived in~\cite{Das:2021fma,Wang:2022jxx}, where the authors computed the photon polarization tensor without making simplifying assumptions or limiting the number of Landau levels. The production of direct photons from a magnetized medium was similarly studied in~\cite{Wang:2020dsr}. In addition, the anisotropy in the emission of photons and dileptons in the presence of such fields was also analyzed in~\cite{Wang:2020dsr,Wang:2022jxx,Wang:2023fst}. Phenomenological estimates of these electromagnetic emissions were further carried out in~\cite{Tuchin:2012mf,Tuchin:2013bda}. Recently,~\cite{Gao:2025prq} employed a $3+1$D hydrodynamic simulation with a phenomenologically decaying magnetic field profile to investigate dilepton spectra and elliptic flow, reporting a low-mass enhancement of $v_2$ with increasing magnetic field strength.

In this article, we mainly focus on the production of dileptons and their anisotropic flow through a proper hydrodynamical approach\textemdash thus taking into account the spacetime evolution of both the field and the fluid, whereas all the above-mentioned previous calculations were static in nature. This is the next logical step which allows us to obtain results directly comparable with experiments. We will utilize the most general expression for the dilepton production rate in a hot, dense, and magnetized medium derived in~\cite{Das:2021fma}. However, for the present analysis we stick to zero baryon density.

The analysis in~\cite{Das:2021fma} revealed a significant enhancement in the dilepton production rate in the low invariant mass region and subsequently led the authors to prescribe a hydrodynamic evolution of their results. Such an enhancement, previously was, either completely absent, or if present, only within a very narrow and nearly undetectable window of invariant mass, in calculations based on limiting cases of the external magnetic field, such as the strong or weak field approximations~\cite{Bandyopadhyay:2016fyd,Bandyopadhyay:2017raf,Das:2019nzv}. In the strong-field regime, often treated using the LLL approximation, this difference arises due to the omission of contributions from quark-antiquark decay processes, which require transitions between multiple Landau levels and are not captured within a single Landau level framework. On the other hand, the weak–field approximation, which also takes only quark-antiquark annihilation processes into account, led to a suppression of the rate before it merged with the Born rate. The enhancement found in~\cite{Das:2021fma} is encouraging, as it can be used as a magnetometer for the fireball created in HIC. Unfortunately, this was a static calculation, and the results cannot be directly compared with the experimental observables. This leads to our present endeavor, where we try to estimate the spectrum using a space-time evolution of the rate~\cite{Das:2021fma}.

To achieve this, we employ a fluid profile derived from the exact hydrodynamic solution incorporating specific symmetries, as introduced in~\cite{Gubser:2010ze,Gubser:2010ui}—commonly referred to as the Gubser solution. The Gubser solution provides a systematic framework that accounts for both the longitudinal expansion of the QGP along the beam axis and the transverse expansion, which becomes dominant in the later stages of the plasma's evolution.

 For the magnetic field profile, we use a realistic configuration corresponding to colliding nuclei at a given impact parameter and finite conductivity, obtained from a semi-analytical solution of Maxwell’s equations for such a setup \cite{Li:2016tel,Siddique:2021smf,Siddique:2022ozg}. This magnetic field is then embedded into the Gubser flow solution to calculate the space–time evolution of the dilepton production rate. Although the Gubser solution possesses azimuthal symmetry in the transverse plane—at odds with the anisotropic nature of the electromagnetic field profile—this approach allows us to disentangle the harmonic contributions originating from the rate itself from those induced purely by the geometry of the colliding nuclei.

It is known that by examining the transverse momentum ($p_T$) distribution, valuable insights can be gained into the thermal properties and collective behavior of the medium~\cite{Chatterjee:2007xk}. Accordingly, in this study we present the \( p_T \) spectra and also flow harmonics. By considering the interplay between the effects due to the external magnetic field and hydrodynamic evolution, our analysis aims to provide insights into how realistic magnetic field profiles can influence the spectra and corresponding flow harmonics. Our results are expected to shed light on possible experimental signatures of magnetic field effects in HICs and contribute to the broader understanding of electromagnetic probes in extreme QCD environments.

This article is organized as follows. In Section~\ref{sec:form}, we introduce the formalism for dilepton emission from a hot, magnetized QGP. Subsection~\ref{subsec:static_rate} presents the static rate calculations and key definitions, while Subsection~\ref{subsec:fluid_dynamics} provides the framework for mapping these static quantities onto a dynamical background. The subsequent Subsection~\ref{ssec:flu_n_mag_prof} introduces the fluid and magnetic field profiles, and Section~\ref{subsec:numerical_setup} describes the numerical setup. In Section~\ref{sec:res}, we present the results, with Subsection~\ref{ssec:spec} devoted to spectra calculations and Subsection~\ref{ssec:elp_fl} to even harmonic flows. A summary of the main findings and conclusions is given in Section~\ref{sec:con}. Finally, two appendices are included: Appendix~\ref{sec:app_born} details the variable transformations and also provides the full expression for the dilepton rate, while Appendix~\ref{sec:cutoff} discusses numerical convergence and cutoff dependence.

\section{Formalism}
\label{sec:form}

In this section, we outline the fundamental ingredients required to compute the dilepton spectra and their anisotropic flow. We begin by presenting the expression for the static dilepton production rate in our system, and then proceed to describe the fluid dynamics and magnetic-field profiles used in the evaluation. We also provide details of the numerical setup used to obtain the results.

\subsection{Static Rate}
\label{subsec:static_rate}
The dilepton production rate per unit spacetime and four-momentum volumes for a hot, arbitrarily magnetized medium can be expressed as the sum of three distinct contributions: quark–antiquark annihilation, quark decay, and antiquark decay. In line with Ref.~\cite{Das:2021fma}, where these individual terms and the total rate have been thoroughly evaluated and analyzed within the imaginary-time formalism, we present the final expressions below:
\begin{align}
    \frac{dN}{d^4xd^4P}\Bigg\vert_{\rm total} = \frac{dN}{d^4xd^4P}\Bigg\vert_{q+\bar{q} \rightarrow \gamma^{*}}+\frac{dN}{d^4xd^4P}\Bigg\vert_{q \rightarrow q + \gamma^{*}}+\frac{dN}{d^4xd^4P}\Bigg\vert_{\bar{q} \rightarrow \bar{q} + \gamma^{*}}, \label{eq:DPR_mag_total}
\end{align}
where, the individual contributions and the associated notations are detailed in the Appendix~\ref{sec:app_born}. Compared to Ref.~\cite{Das:2021fma}, the significant change here is that we have switched from Cartesian position and momentum coordinates to the natural (or rapidity) position and momentum coordinate system. Hence we have the redefined position and momentum space measures as $d^4x = \tau\,d\tau\,d\eta\,dx\,dy$ and $d^4P = M\,dM\,p_T\,dp_T\,dY\,d\phi_p$, where $M$ is the invariant mass, $p_T$ is the transverse momentum, $\phi_p$ is the azimuthal angle in momentum space, $\tau$ is the proper time and $\eta , Y$ are the position and momentum rapidity variables, respectively. These variable transformations are also given in Appendix~\ref{sec:app_born}.\\

At this point we remind the readers that an external magnetic field crucially modifies the kinematic constraints and available dilepton production processes. In particular, only energy and longitudinal momentum are conserved at the quark–photon vertex, and the resulting non-conservation of transverse momentum enables new quark and antiquark decay channels via Landau-level transitions forbidden at $B=0$. These considerations motivate the decomposition in Eq.~\eqref{eq:DPR_mag_total}, where the standard quark–antiquark annihilation channel persists but with kinematics altered by the Landau-level structure. The presence of these new decay modes is a distinctive signature of the magnetized medium and has no counterpart in the limits of vanishing or very strong magnetic fields.

\subsection{Fluid Dynamics}
\label{subsec:fluid_dynamics}
The invariant yield is obtained from the differential production rate in Eq.~\eqref{eq:DPR_mag_total} as  
\begin{equation}\label{eq:yield}
\frac{dN}{d^4P}=\frac{dN}{M \, dM \, p_T \, dp_T \, dY \, d\phi_p} 
= \int d^{4}x \, \frac{dN}{d^{4}x \, d^{4}P}\Theta(T-T_{\mathrm{min}})
= \int \tau \, d\tau \, dx \, dy \, d\eta \; \frac{dN}{d^{4}x \, d^{4}P}\Theta(T-T_{\mathrm{min}}),
\end{equation}
Similarly, the $n$-th order azimuthal flow coefficient is defined as  
\begin{equation}\label{eq:flow}
v_n(M,p_T) =
\dfrac{\displaystyle \int d\phi_p \, \frac{dN}{M \, dM \, p_T \, dp_T \, dY \, d\phi_p} 
\, \cos \big[ n (\phi_p - \Psi_2) \big]}
{\displaystyle \int d\phi_p \, \frac{dN}{M \, dM \, p_T \, dp_T \, dY \, d\phi_p}},
\end{equation}
where \( \Psi_2 \) represents the second-order reaction plane angle, which defines the orientation of measuring the elliptic flow relative to the $x$-axis in the laboratory frame. Since our calculations are already performed in the laboratory frame with the impact parameter aligned along the $x$-axis, we set \( \Psi_2 = 0 \). In this work, we present results only for the harmonics \( n = 2 \) and \( n = 4 \).

Some important points need to be noted here. We present them below:
\begin{itemize}
    \item For the dynamics we assume that the production of dileptons with momentum $P$ at each spacetime point is given by the equilibrium rate of QGP $dN/d^4xd^4P$ from Eq.~\eqref{eq:DPR_mag_total}, which we subsequently use in Eq.~\eqref{eq:yield}. The four-momentum $P^\mu$ can be decomposed into components parallel and orthogonal to the fluid velocity $u^\mu$ (which is now a function of space-time coordinates  $u^\mu(\tau,x,y,\eta)$) i.e.,
 \begin{equation}
     P^\mu=E_p u^\mu + P^{\langle \mu\rangle},
     \label{Eq:LabtoLocal}
 \end{equation}
 where $E_p\equiv P^{\mu} u_{\mu}$ and the orthogonal part is given as
 \begin{align}
  P^{\langle \mu\rangle}\equiv(g^{\mu\nu}-u^\mu u^\nu)P_\nu=\Delta^{\mu\nu}P_{\nu}.
 \end{align}
 Hence, in Eq.\eqref{eq:DPR_mag_total} we replace $p_0\rightarrow E_p$ and $p\rightarrow \sqrt{E_p^2-M^2}$. Since, $\sqrt{P^2}$ is Lorentz invariant it remains intact.

 \item  The expression for the cut-off temperature in Eq.~\eqref{eq:yield} is different from the usual freeze-out scenario valid for the hadrons, which occurs at a single freeze-out temperature $T=T_f$, see for e.g. \cite{Gursoy:2014aka}. Dileptons, on the other hand, are emitted throughout the hydrodynamic evolution until a certain cut-off $T=T_{\mathrm{min}}$ is reached to account for the breakdown of the hydrodynamic description.

 \item In case of an on-shell particle, meaning $\delta(P^2-M_0^2)$, where $M_0$ is the mass of the stable particle with vanishing decay width, we can trivially integrate over $dp_0$
 \begin{equation}
\frac{dN}{d^3 P} =\int dp_0 2p_0\Theta(p^0)\frac{dN}{d^4 P}\delta(P^2-M_0^2).
 \end{equation}
 This expression is also used in some literature, see for e.g.  \cite{Paquet:2023bdx}. 

\end{itemize}

\subsection{Fluid and Magnetic Field Profile}
\label{ssec:flu_n_mag_prof}

\subsubsection{Fluid profile}
\label{sssec:flu_prof}
As noted in the introduction, we employ the analytic solution to the equations of relativistic conformal hydrodynamics obtained by Gubser~\cite{Gubser:2010ze}. This solution describes the boost-invariant longitudinal expansion and the hydrodynamic transverse flow of a circularly symmetric, strongly coupled conformal plasma. The four-velocity \( u^\mu(\tau, \eta, x_\perp) \) is independent of the azimuthal angle \( \phi \). For a detailed account of Gubser’s solution, we refer the reader to Ref.~\cite{Gubser:2010ze} and provide here only a brief summary. The only nonvanishing components of \( u^\mu \) are \( u^\tau \), corresponding to the boost-invariant longitudinal expansion, and \( u^\perp \), which describes the transverse expansion, given as:
\begin{equation}
    u^{\tau} = \cosh \kappa ~; \quad  u^{\bot} = \sinh \kappa,
\label{eq:GubserFlow_velo}
\end{equation}  
where $\kappa$ is the transverse flow rapidity expressed as
 \begin{equation}
    \kappa\left( {x_\bot}, \tau \right) = \text{Arctanh}\!\left(\dfrac{2  q^2  \tau  x_\bot}{1 + q^2  \tau^2 + q^2  x_\bot^2}\right) \;. 
    \label{eq:GubserFlow_kappa}
  \end{equation}
The fluid four-velocity \( u^\mu \) in Gubser’s solution is determined by a single parameter \( q \), which has dimensions of inverse length . The transverse extent of the plasma is inversely proportional to \( q \). The corresponding local energy density of the plasma is then given by~\cite{Gubser:2010ze},
  \begin{align}
    e = \hat{e}_0 \left(\dfrac{4q^2}{\tau}\right)^{4/3}    \left[1 + 2  q^2  \left(\tau^2 + x_\bot^2\right) + q^4  \left(\tau^2 - x_\bot^2\right)^2\right]^{-4/3}\!\!.
    \label{eq:GubserFlowk_en_den}
  \end{align}
The energy density of the plasma, \( e \), is related to the local temperature \( T \) via \( e = g T^4 \), where we take \( g = 11 \), a value consistent with the equation of state of the QGP. The parameters \( q \) and \( \hat{e}_0 \) are chosen to reproduce a realistic final-state spectrum of charged pions and protons at \(\sqrt{s_{NN}} = 200~\mathrm{GeV}\) in the 20--30\% centrality class, which corresponds to impact parameters of approximately 7--8~fm. We find that setting \( \hat{e}_0 = \hat{T}_0^4 \) with \( \hat{T}_0 = 10.6 \) and \( q^{-1} = 6.4~\mathrm{fm} \) yields reasonable agreement with the experimental pion and proton spectra \cite{PHENIX:2003iij}.

\begin{figure}
    \includegraphics[width=0.45\linewidth]{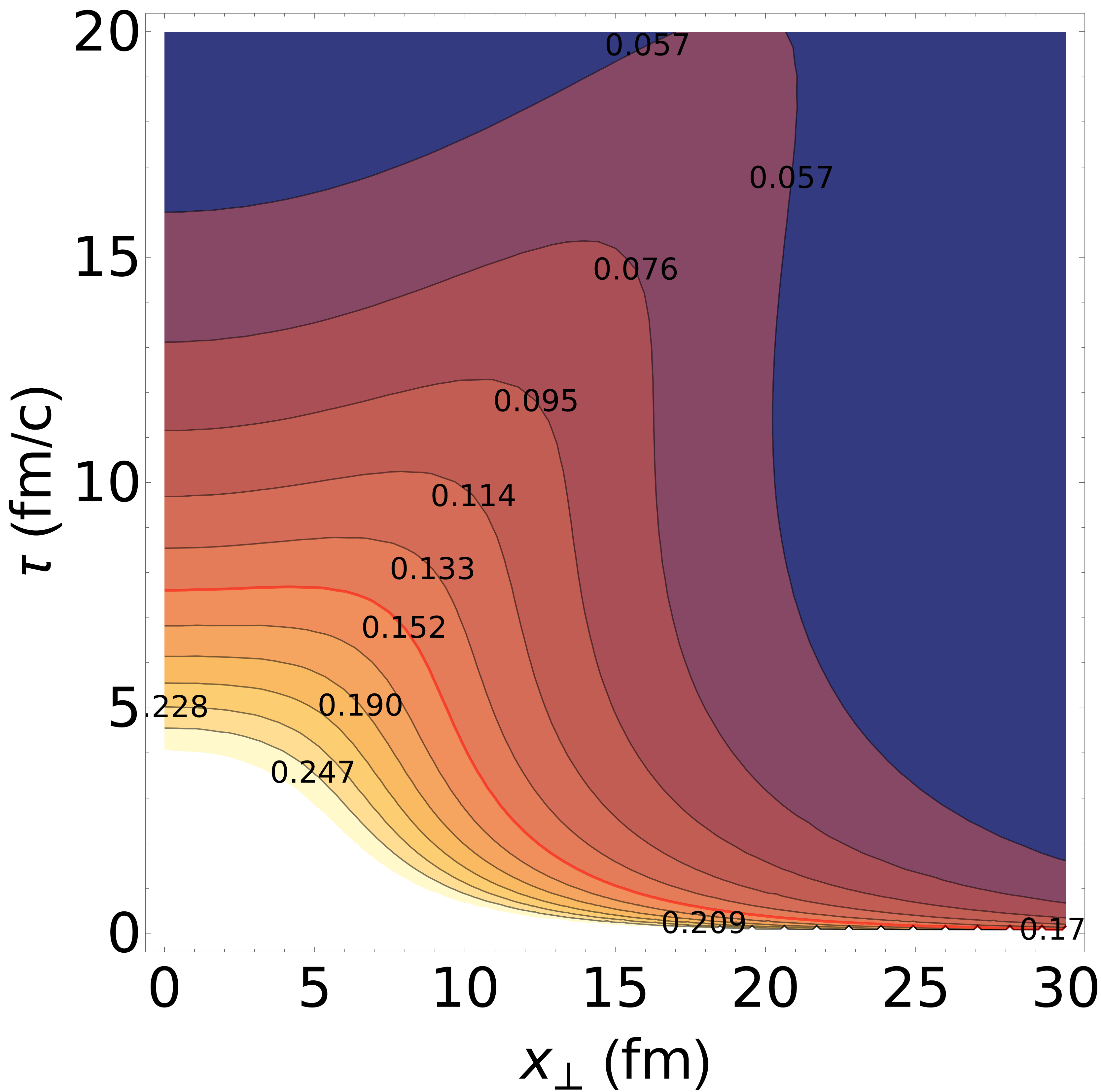}
    \includegraphics[width=0.45\linewidth]{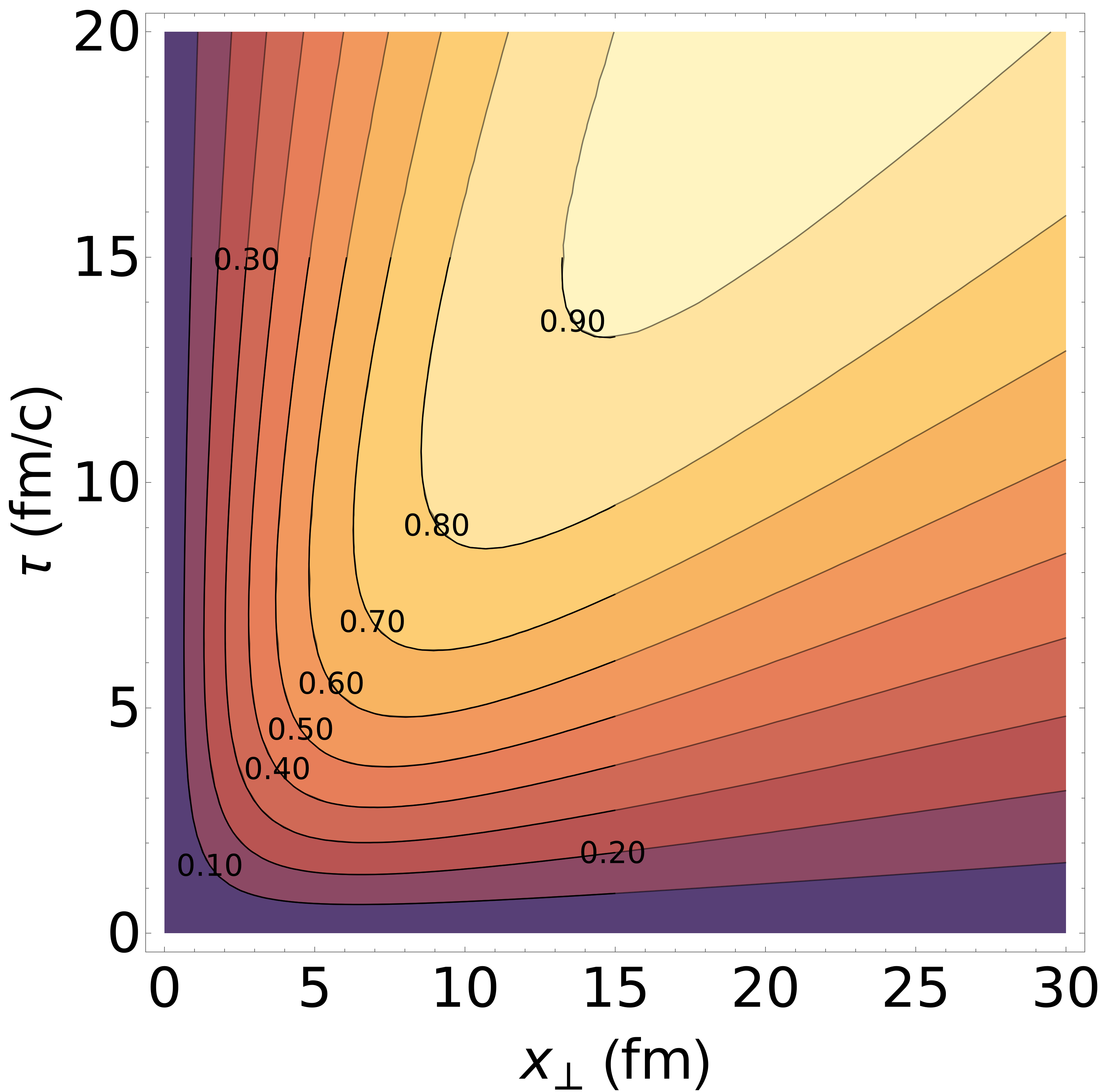}
    \caption{\label{fig:GubserContour} Left panel shows the isothermal contours (in GeV) in the \((x_\perp, \tau)\) plane based on Gubser’s hydrodynamic solution, using the parameters \(\hat{T}_0 = 10.8\) and \(q^{-1} = 6.4\,\mathrm{fm}\). The isotherm (red curve) corresponding to \(T_{\mathrm{min}} = 0.15\,~\mathrm{GeV}\) is selected to represent the freeze-out surface. Right panel shows the contours of transverse velocity $v^\bot(x_\perp, \tau)=u^\bot(x_\perp, \tau)/u^\tau(x_\perp, \tau)$ in the \((x_\perp, \tau)\) plane based on Gubser’s hydrodynamic solution. The parameters are same as the left panel.}
\end{figure}
We define the isothermal surface at \( T_\mathrm{min} = 0.15~\mathrm{GeV} \) (cf. Eq.~\eqref{eq:yield}) to represent the breakdown of hydrodynamics and the suppression of dilepton emission, beyond which the QGP hadronizes. The isothermal contours, including the \( T_\mathrm{min} \) hypersurface, are shown in the left panel of Fig.~\ref{fig:GubserContour}. The right panel of Fig.~\ref{fig:GubserContour} displays the contour plot of the transverse velocity in the \( (x_\perp, \tau) \) plane, based on the above parameter set.

\begin{figure}
    \centering
    \includegraphics[width=0.5\linewidth]{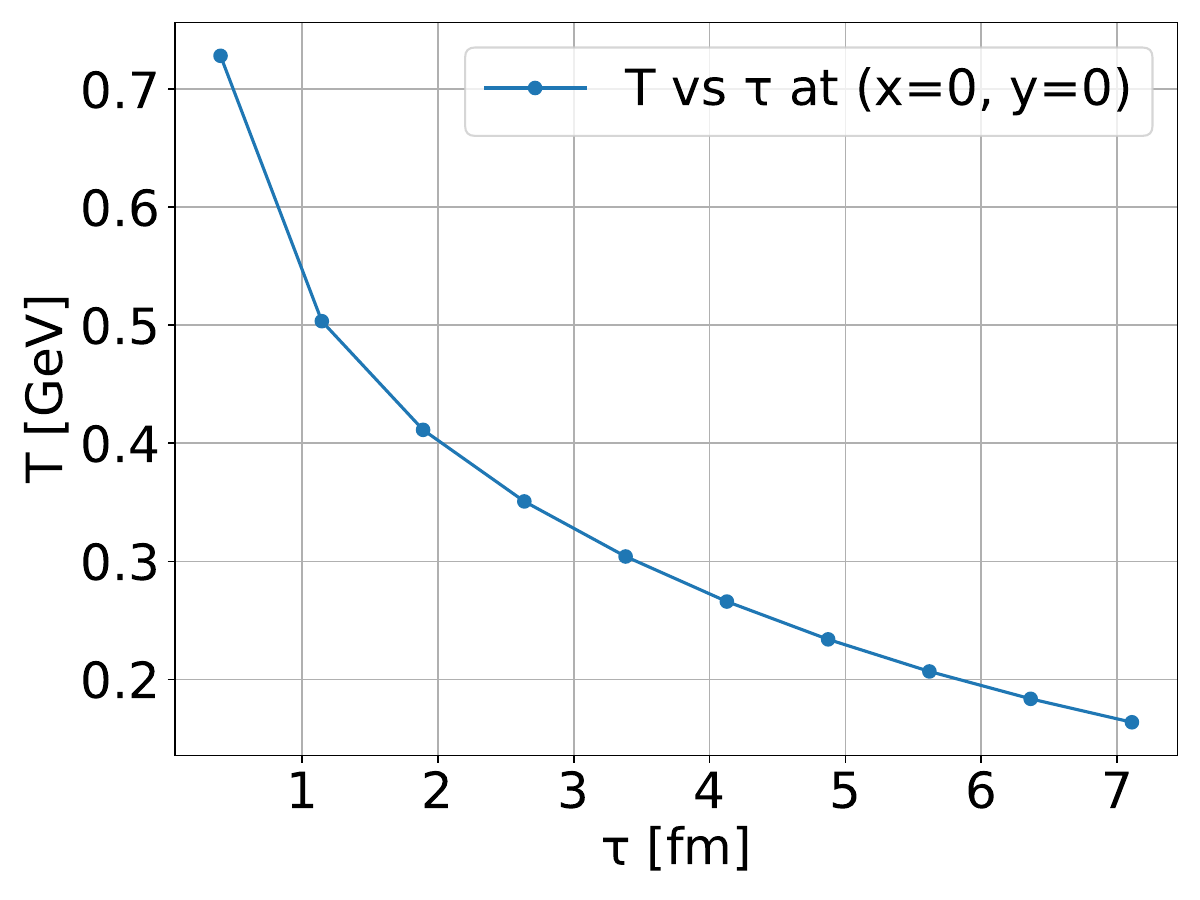}
    \caption{Temperature evolution with proper time at the center of the collision zone $(x, y) = (0, 0)$ for Au+Au collisions.}
     \label{fig:TvsTau}
\end{figure}

With Gubser flow profile with the fluid profile starting at $\tau$ = 0.4 fm, we get the temperature vs proper time plot which is shown in Fig.~\ref{fig:TvsTau}. We then embed this hydrodynamic background in the magnetic field derived in the following section to evaluate the resulting small, charge-dependent correction to the flow.

\subsubsection{Magnetic field profile}
\label{sssec:mag_prof}
Before setting up the numerics to determine the effect of magnetic fields on the dilepton spectra and anisotropic flow, we review the electromagnetic field configurations employed in the current analysis. We begin by presenting the formula for all components of the fields, assuming finite and constant electrical and chiral conductivities that are used in this study. Several works~\cite{Gursoy:2014aka,Siddique:2021smf,Siddique:2022ozg} have previously studied these fields extensively, which are evaluated by considering that all source charges propagate along the $z$-axis and by employing the Green's function method in cylindrical coordinates~\cite{Li:2016tel,Siddique:2021smf,Siddique:2022ozg}. The expressions for the electromagnetic fields for charge $Q$ are as follows:
\begin{align}
B_{\phi}(t,\textbf{x}) &= \frac{Q}{4\pi} \frac{v\gamma x_{\bot}}{\Delta^{3/2}} 
\left(1+\frac{\sigma v\gamma}{2} \sqrt{\Delta} \right) e^{A}, \nonumber \\
B_{r}(t,\textbf{x}) &= -\sigma_{\chi} \frac{Q}{8\pi} \frac{v\gamma^{2}x_{\bot}}{\Delta^{3/2}} e^{A} 
\left[\gamma (vt-z) + A\sqrt{\Delta} \right],\quad {\rm and} \nonumber \\
B_{z}(t,\textbf{x}) &= \sigma_{\chi} \frac{Q}{8\pi} \frac{v\gamma}{\Delta^{3/2}} e^{A} 
\left[\Delta \left(1 - \frac{\sigma v\gamma}{2} \sqrt{\Delta} \right) 
+ \gamma^{2} (vt-z)^{2} \left(1 + \frac{\sigma v\gamma}{2} \sqrt{\Delta} \right) \right], 
\label{eq:eq8}
\end{align}
where,
\begin{align}
\Delta &= \gamma^{2} (vt - z)^{2} + x_{\bot}^{2}, \quad {\rm and} \\
A &= \frac{\sigma v\gamma}{2} \left[\gamma (vt - z) - \sqrt{\Delta} \right],
\end{align}
with $x_{\bot}$ representing the transverse coordinate magnitude, $x_{\bot} = \sqrt{x^{2} + y^{2}}.$

With the average nuclear charge density $\bar{\rho} = Z/(\frac{4}{3} \pi R^3)$, where $Z$ is the atomic number of colliding nuclei,  the magnetic fields of the projectile nucleus moving in $+z$-direction and the one of the target nucleus moving in $-z$-direction are evaluated as:
\begin{widetext}
 \begin{eqnarray}
   \boldsymbol{B}_{\text{proj}}\!\left(x_-, \vert \boldsymbol{b}_1\vert \right) &=& 
     \int \mathrm{d}^2 b^{\prime}  \, 2\, \bar{\rho} \, \sqrt{R^2 - {b^{\prime\,2}}} \,\, \boldsymbol{B}\!\left(x_-, \vert \boldsymbol{b}_1 - \boldsymbol{b}^{\prime} \vert\right) \cdot 
     \left(-\sin\psi_1 \boldsymbol{e}_x + \cos\psi_1 \boldsymbol{e}_y\right) 
     \;,\label{Eq:Bproj} \\
   \boldsymbol{B}_{\text{targ}}\!\left(x_+, \vert \boldsymbol{b}_2\vert \right) &=& 
     \int \mathrm{d}^2 b^{\prime}  \, 2\, \bar{\rho} \, \sqrt{R^2 - {b^{\prime\,2}}} \,\, \boldsymbol{B}\!\left(x_+, \vert \boldsymbol{b}_2 - \boldsymbol{b}^{\prime} \vert\right) \cdot 
     \left(-\sin \psi_2 \boldsymbol{e}_x + \cos\psi_2 \boldsymbol{e}_y\right) 
     \;\label{Eq:Btarg},
 \end{eqnarray}
\end{widetext}
where $x_{\mp} := t \mp z/v$, while $\psi_{\{1,2\}}$\footnote{Both projectile and target components are presented by a compact notation.} is the angle between the vector $\boldsymbol{b}_{\{1,2\}} - \boldsymbol{b}^{\,\prime}$ and the $x$-axis, 
 \begin{equation}
   \cos \psi_{\{1,2\}} = 
     \dfrac{b_{\{1,2\}} \cos \phi_{\{1,2\}} - b^{\prime} \cos\phi^{\prime}}
     {\sqrt{b_{\{1,2\}}^2 + b^{\prime \, 2} - 2  \, b_{\{1,2\}} \, b^{\prime} \cos\!\left(\phi^{\prime} - \phi_{\{1,2\}}\right)}}  \;.
 \end{equation}
We note that the total magnetic field is then simply the sum of the two, $\boldsymbol{B}_{\text{tot}} = \boldsymbol{B}_{\text{proj}} + \boldsymbol{B}_{\text{targ}}$, and while the transformation from cylindrical to Cartesian coordinates is straightforward, the transformation from Cartesian to Milne coordinates reads as~\cite{Inghirami:2019mkc}:
  \begin{equation}
    \tilde{B}_x = \dfrac{B_x}{\cosh \eta} \;,
  \quad
    \tilde{B}_y = \dfrac{B_y}{\cosh \eta} \;,
  \quad
    B_{\eta}  = \dfrac{B_z}{\tau} \;.
  \end{equation}

One would like to know the electromagnetic field at a given impact parameter \( \mathbf{b} = \mathbf{b}_1 - \mathbf{b}_2 \) as a function of time \( t \) and spatial coordinates \( (x, y, z) \), such that the $x$-axis being the direction of impact parameter and the $z$-axis being the direction of collision, the angles and magnitudes are given by:
\[
\tan \phi_{\{1,2\}} = \frac{y}{x \pm \frac{b}{2}}, \quad 
b_{\{1,2\}} = \sqrt{\left(x \pm \frac{b}{2}\right)^2 + y^2}
\]
Although, as shown in \cite{Dash:2023kvr}, the contribution of participants to the electromagnetic field is non-negligible, it is 
neglected in this work for the sake of simplicity. The chiral conductivity is set to zero throughout the present analysis. Lattice-QCD calculations predict that the electric conductivity of a static QGP is relatively large \cite{Li:2016tel, Pasechnik:2016wkt},  
\begin{equation}
    \sigma = \left(5.8 \pm 2.9\right) \frac{T}{T_C} \ \mathrm{MeV},
\end{equation}
where \( T \) is the plasma temperature and \( T_C \) is the transition temperature between the QGP and the hadronic phase. Although the conductivity is, in general, temperature-dependent, we assume a uniform value for simplicity. To facilitate comparison, we consider three representative conductivities: \( 0.58 \ \mathrm{MeV} \), \( 5.8 \ \mathrm{MeV} \), and \( 10.6 \ \mathrm{MeV} \) throughout the analysis.
 
\begin{figure}
            \centering
            \includegraphics[width=\linewidth]{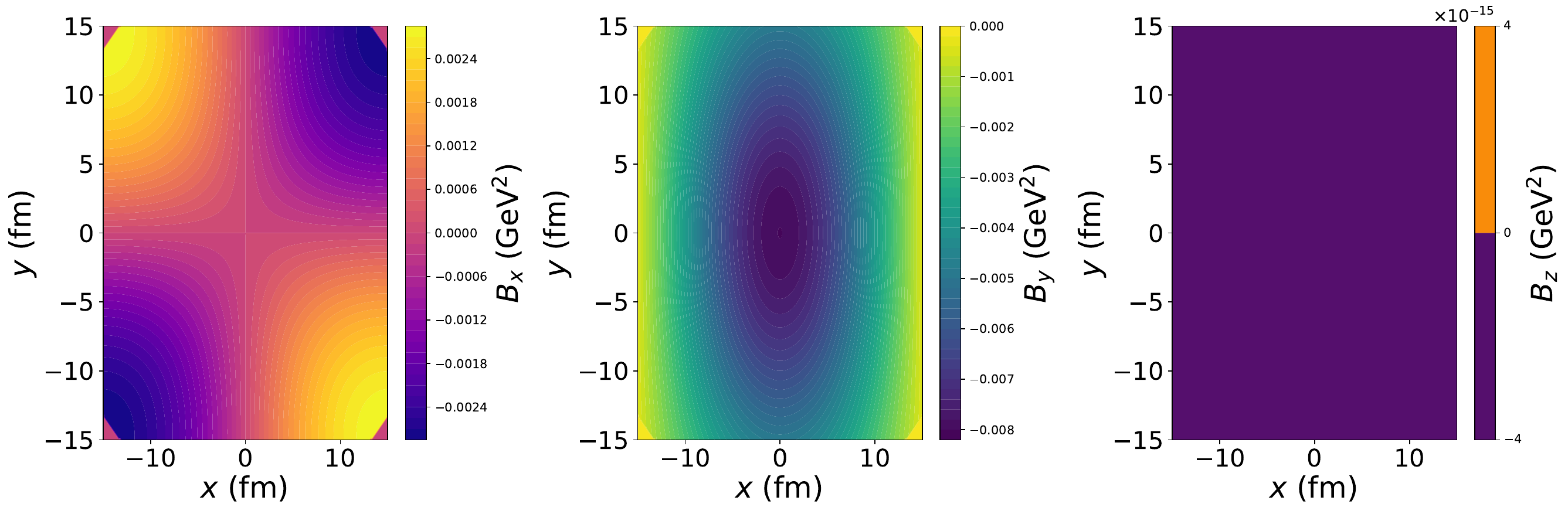}
            \caption{Transverse profiles of different components of the magnetic field for Au+Au collisions at \( \sqrt{s_{NN}} = 200 \)~GeV with impact parameter \( b = 12 \)~fm, evaluated at proper time \( \tau = 0.4 \)~fm and electrical conductivity \( \sigma = 0.58 \)~MeV.}
            \label{fig:mag_prof}
\end{figure}

 The transverse profiles of different components of the magnetic field for Au+Au collisions at \( \sqrt{s_{NN}} = 200 \)~GeV, with impact parameter \( b = 12 \)~fm, evaluated at proper time \( \tau = 0.4 \)~fm and electrical conductivity \( \sigma = 0.58 \)~MeV, are shown in Fig.~\ref{fig:mag_prof}. From the figure, it is clear that the \( y \)-component of the magnetic field dominates in the central collision zone, whereas the \( x \)-component becomes more prominent in the peripheral region. This behavior is consistent with previously reported results~\cite{Deng:2012pc}.

\subsection{Numerical Setup}
\label{subsec:numerical_setup}
For the numerical setup, we employed a \((2+1)\)-dimensional Milne grid covering the domain \([-15~\mathrm{fm}, 15~\mathrm{fm}] \times [-15~\mathrm{fm}, 15~\mathrm{fm}] \times [0.4~\mathrm{fm}/c, 6.0~\mathrm{fm}/c]\), discretized into \(200 \times 200 \times 40\) cells along the \((x, y, \tau)\) directions in space-time. The fluid velocity and energy density (or temperature) at each grid point were computed using Eqs.~\eqref{eq:GubserFlow_velo} and~\eqref{eq:GubserFlowk_en_den}, respectively. Only those space-time regions satisfying \(T(x, y, \tau) > T_{\mathrm{min}}\) were considered in the analysis.

The electromagnetic fields at these points were embedded using Eqs.~\eqref{Eq:Bproj} and~\eqref{Eq:Btarg}, for various impact parameters and values of electrical conductivity. Since the dilepton production rate is computed under the assumption of a constant magnetic field aligned along a given coordinate axis, we retained only the dominant component, \(B_y\), in our calculations, while all other components were set to zero.

The local values of temperature, fluid velocity, and magnetic field strength \(B_y\) were then used for the numerical evaluation of Eqs.~\eqref{eq:rate_ann},~\eqref{eq:rate_decay_p}, and~\eqref{eq:rate_decay_ap}, making use of the transformation relations in Eq.~\eqref{Eq:LabtoLocal} to convert lab-frame four-momenta to the local rest frame in each fluid cell.

\begin{figure}
    \centering
    \includegraphics[width=0.45\textwidth]{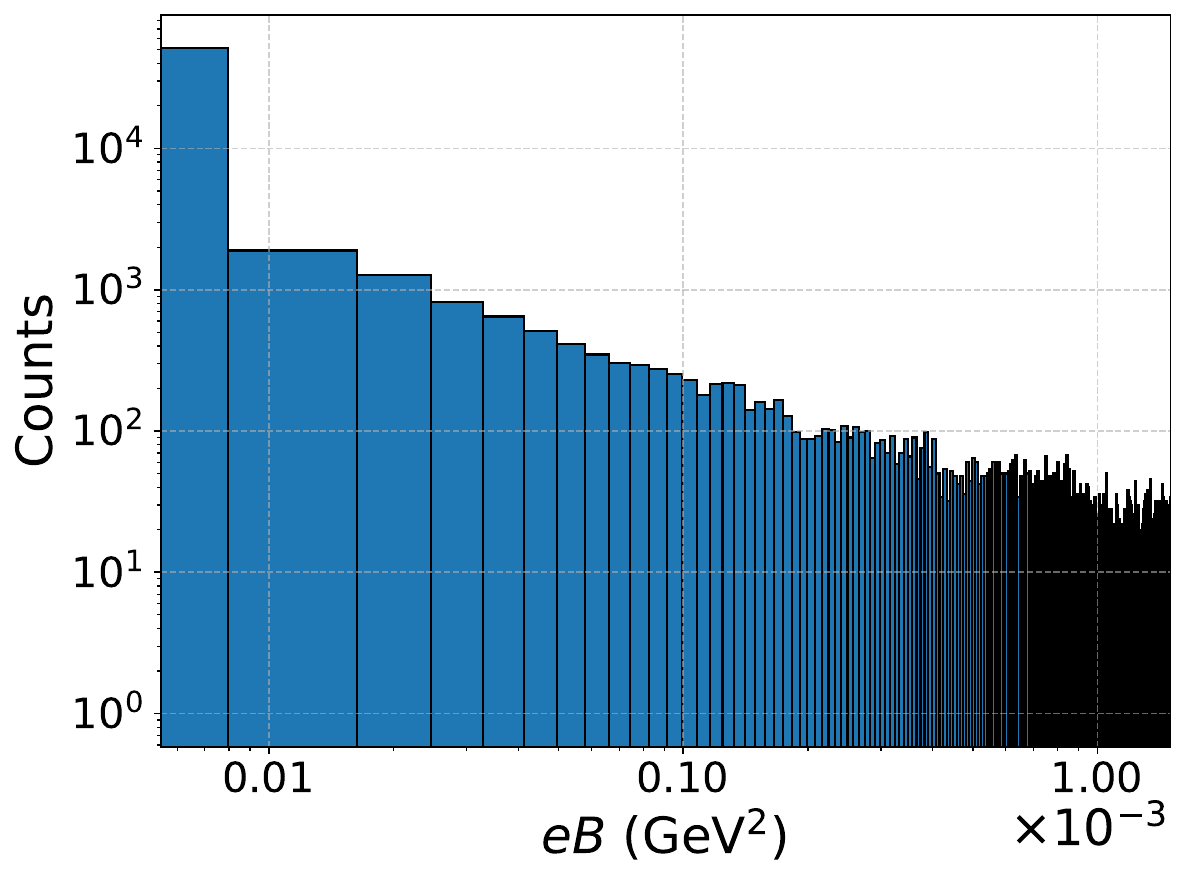}
    \includegraphics[width=0.45\linewidth]{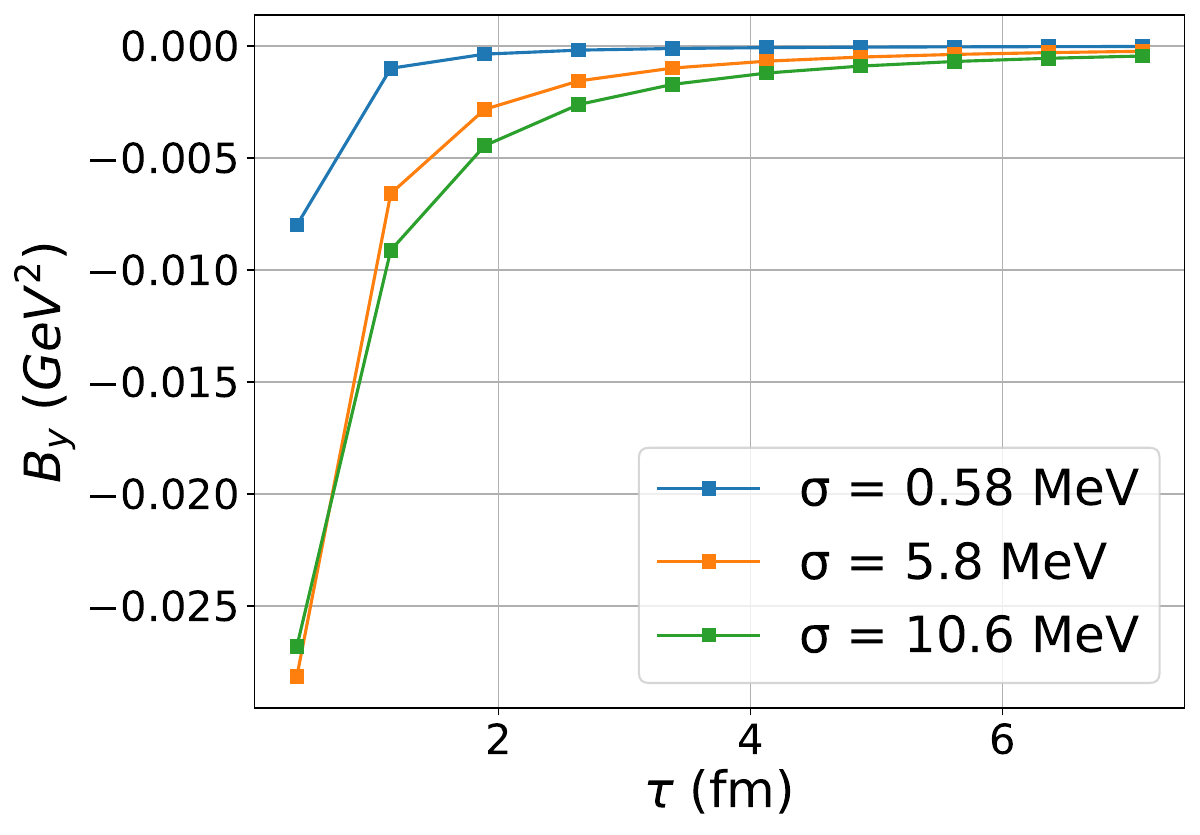}
    \caption{Left panel: histogram showing the distribution of the combined magnetic field components \( eB \in \{eB_x, eB_y, eB_z\} \) for \( b = 12 \)~fm and \( \sigma = 5.8 \)~MeV. Right panel: time evolution of \( B_y \) for different values of electrical conductivity at \( \sqrt{s_{\mathrm{NN}}} = 200 \)~GeV and impact parameter \( b = 12 \)~fm, evaluated at the spatial point \( (x, y) = (0, 0) \) for Au+Au collisions.}
    \label{fig:Bfield_histogram}
\end{figure}

The left panel in Fig.~\ref{fig:Bfield_histogram} shows the histogram of magnetic field  for $b=12$ fm and $\sigma=5.8$ MeV across the whole hyper-volume of fireball evolution, while the Fig.~\ref{fig:Bfield_histogram} (right panel) shows the temporal variation of magnetic field across the above setup for various conductivities. As can be seen from the figure the magnitude of magnetic field peaks around $\simeq10^{-5}~\mathrm{GeV}^2$ for most of the fire ball region. The mean value of the magnetic field $\langle B_y\rangle$ across all fluid cells, is around $\simeq10^{-4}~\mathrm{GeV}^2$.  

The dilepton production rates in Eqs.~\eqref{eq:rate_ann}, \eqref{eq:rate_decay_p}, and~\eqref{eq:rate_decay_ap} involve generalized Laguerre polynomials \( L_{\ell}^{\alpha}(\xi) \), where \( \xi = {p_{\perp}^2}/(2|q_{\mathrm{f}}B|) \) with $p_\perp$ being the perpendicular component of the momentum $p$ with respect to the magnetic field direction, defined in Eq.~\eqref{eq:pl_n_pp}. When the magnetic field strength in a given fluid cell is too small, the argument \(\xi\) of the Laguerre polynomial becomes large, which can lead to numerically unstable or imprecise evaluations due to the large magnitude of the polynomials. To mitigate this issue, we introduce a cutoff \(\xi_{\mathrm{cf}}\) on the argument \(\xi\), which depends on the invariant mass \(M\) (since \(p_\perp\) depends on \(M\)). For \(M \in \{0.1,\, 0.3,\, 0.5\}\)~GeV, we set \(\lfloor \xi_{\mathrm{cf}} \rfloor \in \{120,\, 100,\, 50\}\), respectively. Similarly, we limit the number of Landau levels summed over in the case of very weak magnetic fields by introducing a cutoff on the magnetic field strength, requiring \(|eB| > 0.003~\mathrm{GeV}^2\). When the above conditions are not satisfied for a given fluid cell, we revert to the Born rate expression for the annihilation channel, while assigning zero contribution from the decay processes. This is justified since the annihilation rate approaches the Born rate in the limit of a weak magnetic field.

\section{Results}
\label{sec:res}
In this section we explore the effects of the strength and lifetime of the magnetic field on the dilepton spectra and anisotropic flow. The magnetic field strength is controlled by varying the impact parameter $b$, while the lifetime of the magnetic field is modified by changing the conductivity $\sigma$ of the medium. Additionally, results are presented for various values of the dilepton invariant mass $M$. Whenever we vary the setup across different $b$ or $M$, we consider the fixed value $\sigma = 5.8~\mathrm{MeV}$, which corresponds to the LQCD value at $T = T_C$. Conversely, when varying the configuration across different $\sigma$ or $M$, we fix $b = 12~\mathrm{fm}$, which corresponds to the peak strength of the magnetic field generated in HICs at $\sqrt{s} = 200~\mathrm{GeV}$~\cite{Deng:2012pc,Panda:2024ccj}. For all the variations with respect to $b$ or $\sigma$, we set $M = 0.1~\mathrm{GeV}$, since the effects of the magnetic field are most pronounced in the region of low dilepton invariant mass.

\subsection{Spectra}
\label{ssec:spec}

\subsubsection{The effect of the strength of the magnetic field}
\label{sssec:eff_str_spec}
\begin{figure}[htbp]
    \centering
    \begin{subfigure}{0.325\textwidth}
        \includegraphics[width=\textwidth]{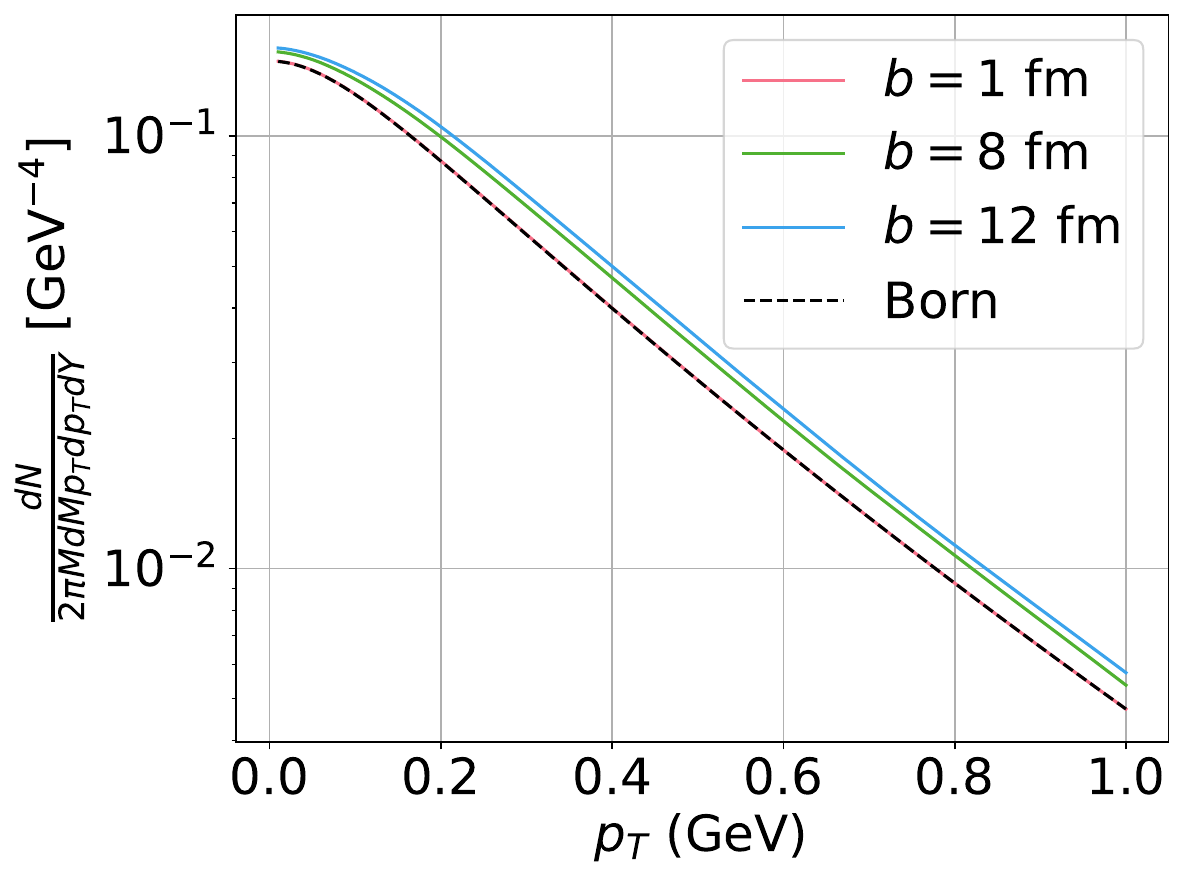}
        \label{fig:spectra_b_total}
    \end{subfigure}
    \hfill
    \begin{subfigure}{0.325\textwidth}
        \includegraphics[width=\textwidth]{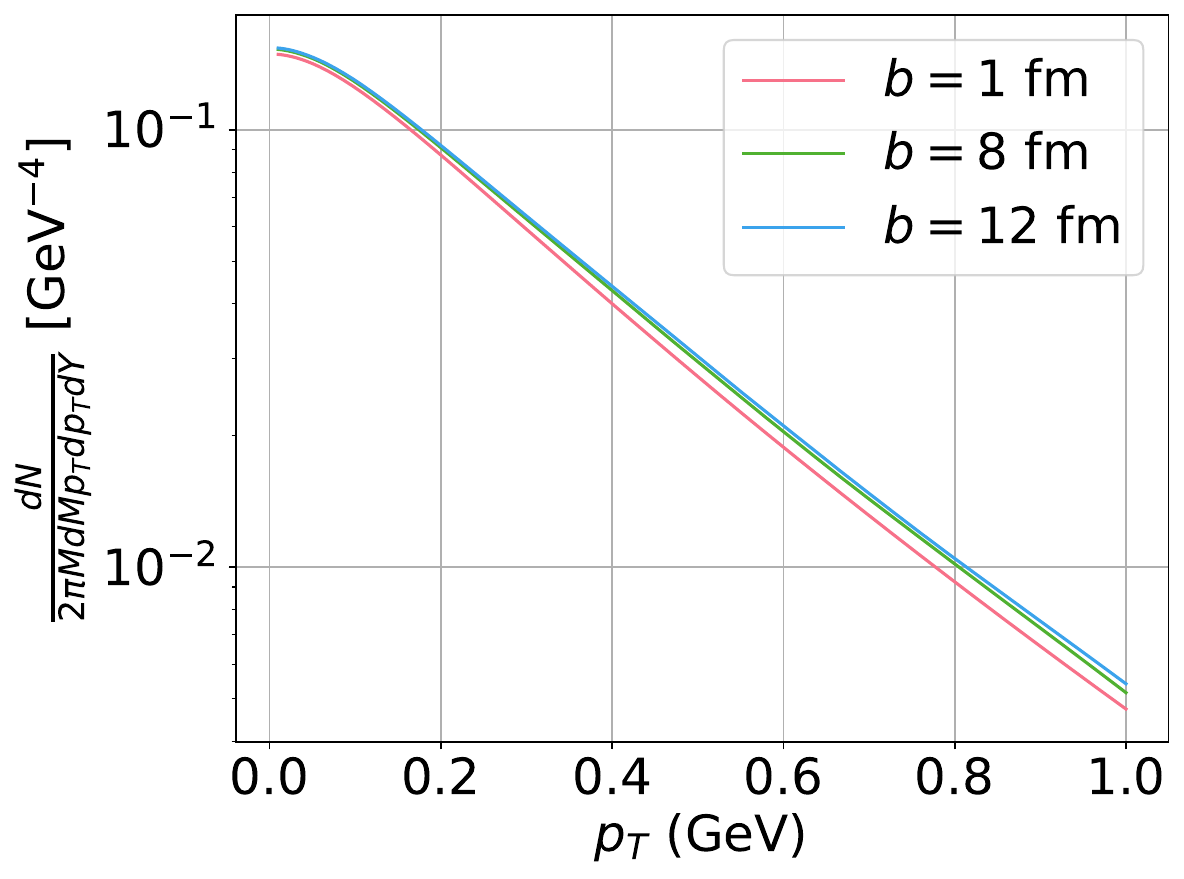}
        \label{fig:spectra_b_anni}
    \end{subfigure}
    \hfill
    \begin{subfigure}{0.325\textwidth}
        \includegraphics[width=\textwidth]{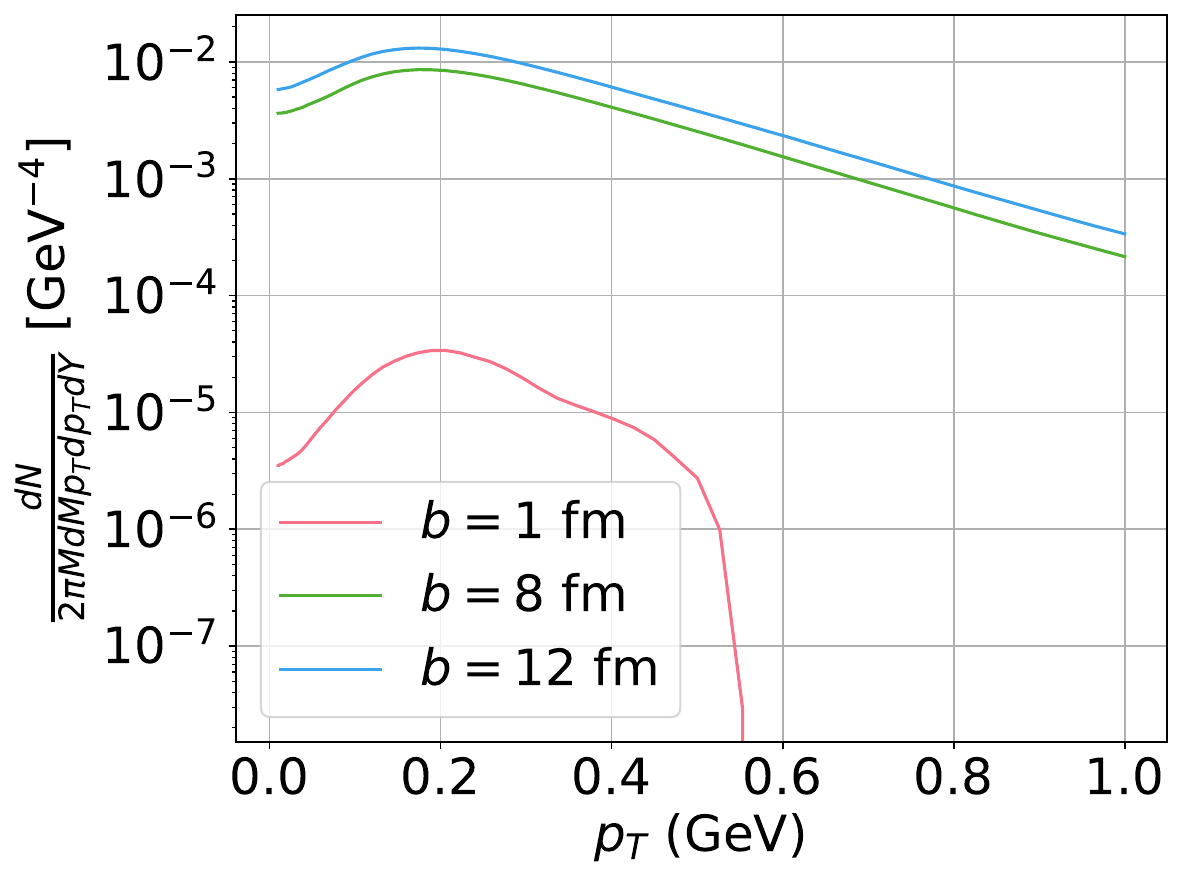}
        \label{fig:spectra_b_decay}
    \end{subfigure}
    \caption{Transverse momentum spectra $dN/(2\pi M dM p_T dp_T dY)$ for various impact parameters at fixed ($M$, $\sigma$, $Y$) = (0.1 GeV, 5.8 MeV, 0), showing the total (left) and the contribution of annihilation (middle) and decay (right) processes. The spectra corresponding to the Born rate (dotted line) has also been indicated in the total spectra plot.}
    \label{fig:spectra_comparison}
\end{figure}
In Fig.~\ref{fig:spectra_comparison}, we present the transverse momentum ($p_T$) spectra (scaled by a $2\pi$ factor) of dileptons at mid-rapidity for various impact parameters. Although the total spectrum is the observable of primary interest, we decompose it into individual processes to understand their relative contributions. The left panel shows the total yield, the middle panel corresponds to dileptons produced via the annihilation process, and the right panel depicts those produced through particle/antiparticle decay. The Born rate (dotted line) has also been shown in the left panel. In both processes, the spectra increase with increasing magnetic field strength, with the enhancement attributed to the background field. For annihilation, the enhancement becomes more pronounced at higher $p_T$, whereas for decay, it is present across the entire $p_T$ range.

It is noteworthy that the yield from annihilation is significantly larger in magnitude, indicating the dominance of this process over decay in dilepton production. In the absence of a magnetic field, this is the only surviving contribution, as seen from the $b = 1~\mathrm{fm}$ curves, which coincide with the Born rate. Thus, the dilepton yield at zero field is essentially governed by the Born contribution. Consequently, the presence of a background magnetic field leads to an enhancement of the total $p_T$ spectrum, as shown in the left panel of Fig.~\ref{fig:spectra_comparison}. This enhancement persists across the full $p_T$ range considered and becomes more pronounced with increasing magnetic field strength. The effect of the cutoff parameter $\xi_{\mathrm{cf}}$ on the overall spectra is discussed in Appendix~\ref{sec:cutoff}.

Another interesting feature in the spectra shown above, is the absence of the sharp spikes typically seen in electromagnetic emission calculations, which arise from Landau level dependent kinematic thresholds~\cite{Wang:2020dsr,Wang:2021ebh,Das:2021fma,Wang:2022jxx}. Reference~\cite{Wang:2021ebh} argued that particle interaction effects could potentially smear out these spikes. In the present calculation, however, the spikes are smoothed out due to the superposition of contributions from the Landau level sum across different space-time points. For the same reason, we do not observe any spikes in the flow harmonics either, as will be shown in subsection \ref{ssec:elp_fl}.

\subsubsection{The effect of the lifetime of the magnetic field}
\label{sssec:eff_life_spec}
\begin{figure}[htbp]
    \centering
    \includegraphics[scale=0.4]{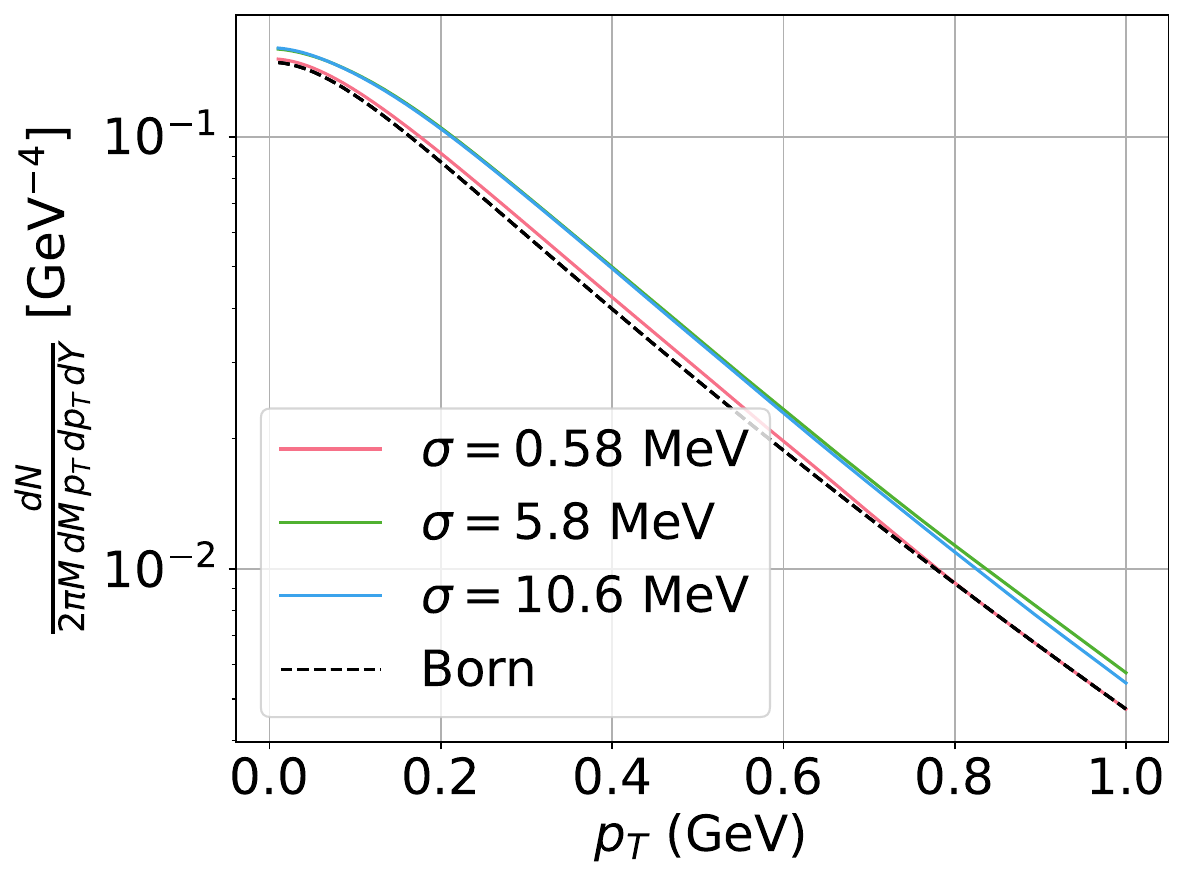}    
    \caption{Total transverse momentum spectra $dN/(2\pi M dM p_T dp_T dY)$ for different electrical conductivities at given ($M$, $b$, $Y$) = (0.1 GeV, 12 fm, 0). The spectra corresponding to the} Born rate (dotted line) has also been indicated.
    \label{fig:spectra_comparison_conductivities}
\end{figure}
Before discussing the influence on the spectra, we note that for finite conductivity, the expression in Eq.~\eqref{eq:eq8} exhibits some non-trivial behavior, particularly when comparing the solutions at finite conductivity to the vacuum solution, which corresponds to the Liénard–Wiechert potential. One observes a strong suppression of the finite-conductivity solution relative to the vacuum case. This suppression arises from the specific form of the solution, which contains an exponential factor $\sim \exp(-A)$. Consequently, the initial electromagnetic fields at early times are smaller for finite conductivity than in the vacuum case, and the suppression becomes stronger with increasing conductivity. This behavior is illustrated in the right panel of Fig.~\ref{fig:Bfield_histogram}, which shows the field lifetime for different values of~$\sigma$. Although the suppression relative to the vacuum is not directly visible in the figure (since our simulation begins at $\tau \sim 0.4$), it is clearly observed at early times when comparing the two larger finite conductivity cases. This behavior contrasts sharply with numerical solutions of Maxwell's equations~\cite{McLerran:2013hla,Dash:2022xkz}, which do not display such suppression. 

We employ the above analytic expression to simplify the electromagnetic field calculation; however, the results should be interpreted with caution. Although these solutions capture the late-time behavior qualitatively, the magnetic field strength at late times becomes too small to have any substantial effect on our analysis.  In Fig.~\ref{fig:spectra_comparison_conductivities}, we explore the effect of the external field lifetime on the total spectra by varying the strengths of $\sigma$.

From Fig.~\ref{fig:spectra_comparison_conductivities}, one observes an increase in the particle yield in the presence of finite conductivity compared to the vacuum case (which would roughly correspond to $\sigma = 0.58~\mathrm{MeV}$). This behavior is expected, since finite conductivity allows the magnetic field to persist for a longer lifetime, leading to a larger average magnetic field in the space–time volume. However, when comparing the latter two finite-conductivity cases, there is a non-trivial depletion of particles for the larger conductivity relative to the smaller one, at high $p_T$. As discussed earlier, this is a feature of the model rather than a genuine physical effect. As expected, the vacuum case also nearly coincides with the Born rate, showing only a slight enhancement at intermediate $p_T$.

\subsubsection{The effect of the invariant mass}
\label{sssec:eff_mass_spec}
\begin{figure}[htbp]
    \centering
    \includegraphics[scale=0.4]{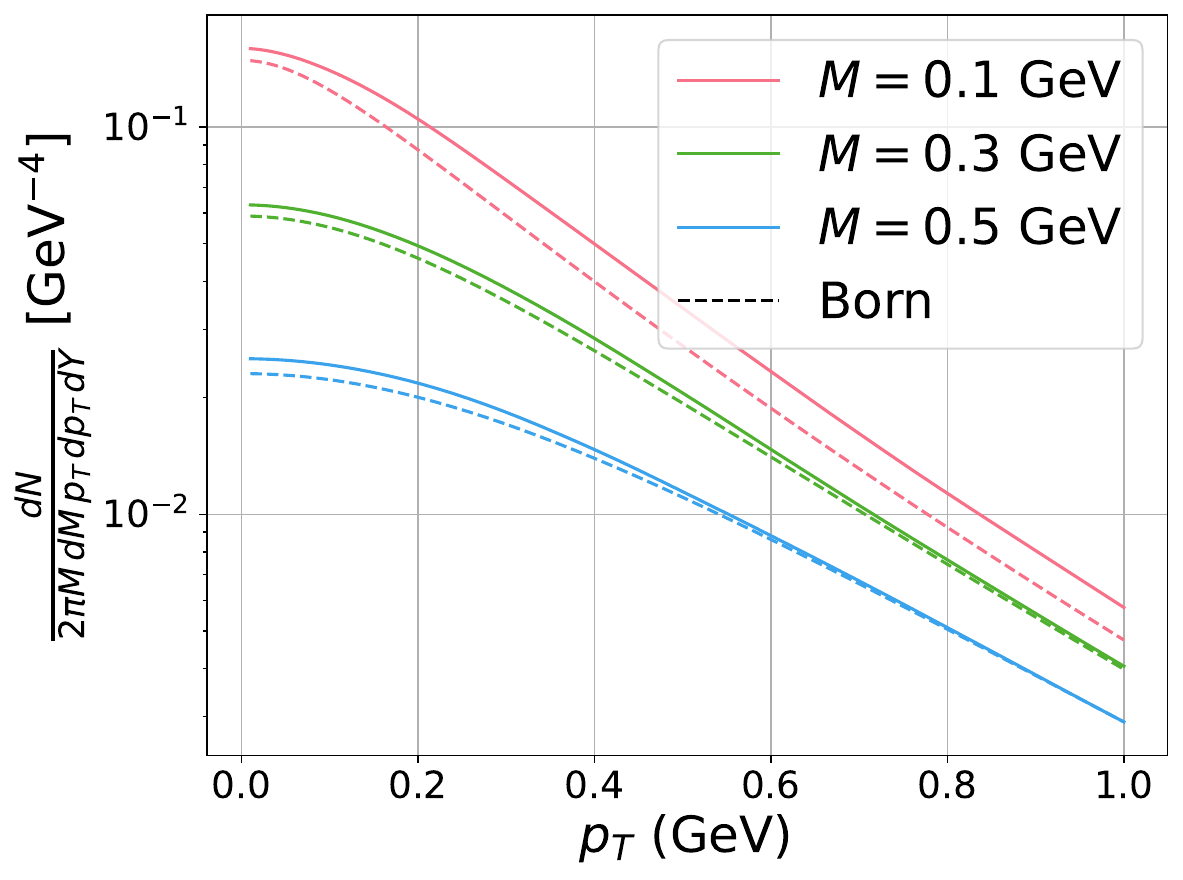}    
    \caption{Total transverse momentum spectra $dN/(2\pi M dM p_T dp_T dY)$ for different invariant masses at given ($b$, $\sigma$, $Y$) = ($12~{\rm fm}, 5.8~{\rm MeV}, 0$). The spectra corresponding to the Born rate (dotted line) has also been indicated.}
    \label{fig:spectra_comparison_mass}
\end{figure}

Fig.~\ref{fig:spectra_comparison_mass}, shows the impact of the invariant mass on the total spectra.  We display results at midrapidity for three values of $M = (0.1$, $0.3$, $0.5)\,\,{\rm GeV}$, along with the respective Born spectrum.  A clear enhancement in the dilepton yield due to the background magnetic field is observed for all invariant masses considered. This enhancement extends to higher $p_T$ values when the invariant mass is small, but gradually shifts to lower $p_T$ as the invariant mass increases. Overall, the dependence on invariant mass is straightforward and consistent with expectations: the spectra are increasingly suppressed as the invariant mass grows.

\subsection{Even Flow Harmonics}
\label{ssec:elp_fl}

\subsubsection{The effect of the strength of the magnetic field}
\label{sssec:eff_str_flow}
\begin{figure}[htbp]
    \centering
    \begin{subfigure}{0.32\textwidth}
        \includegraphics[width=\textwidth]{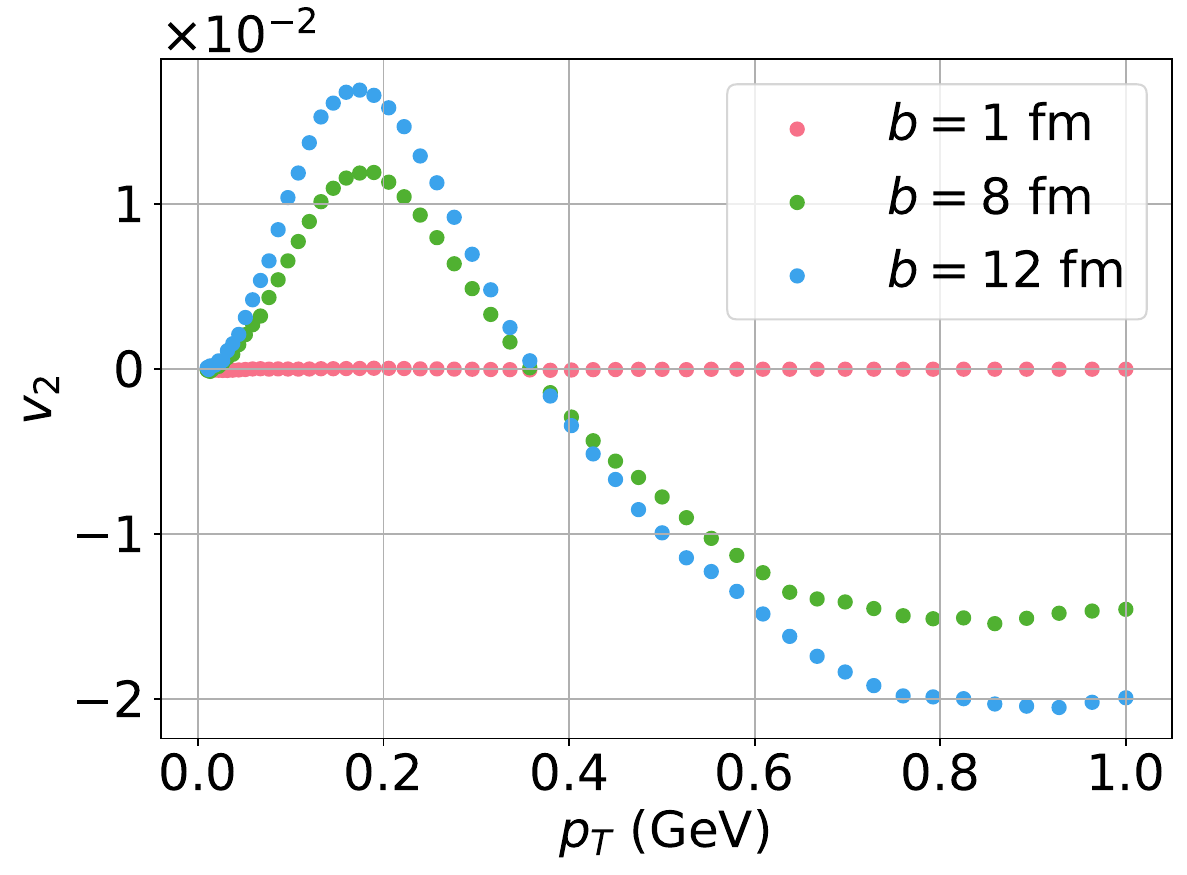}
        \label{fig:v2_b_total}
    \end{subfigure}
    \hfill
    \begin{subfigure}{0.32\textwidth}
        \includegraphics[width=\textwidth]{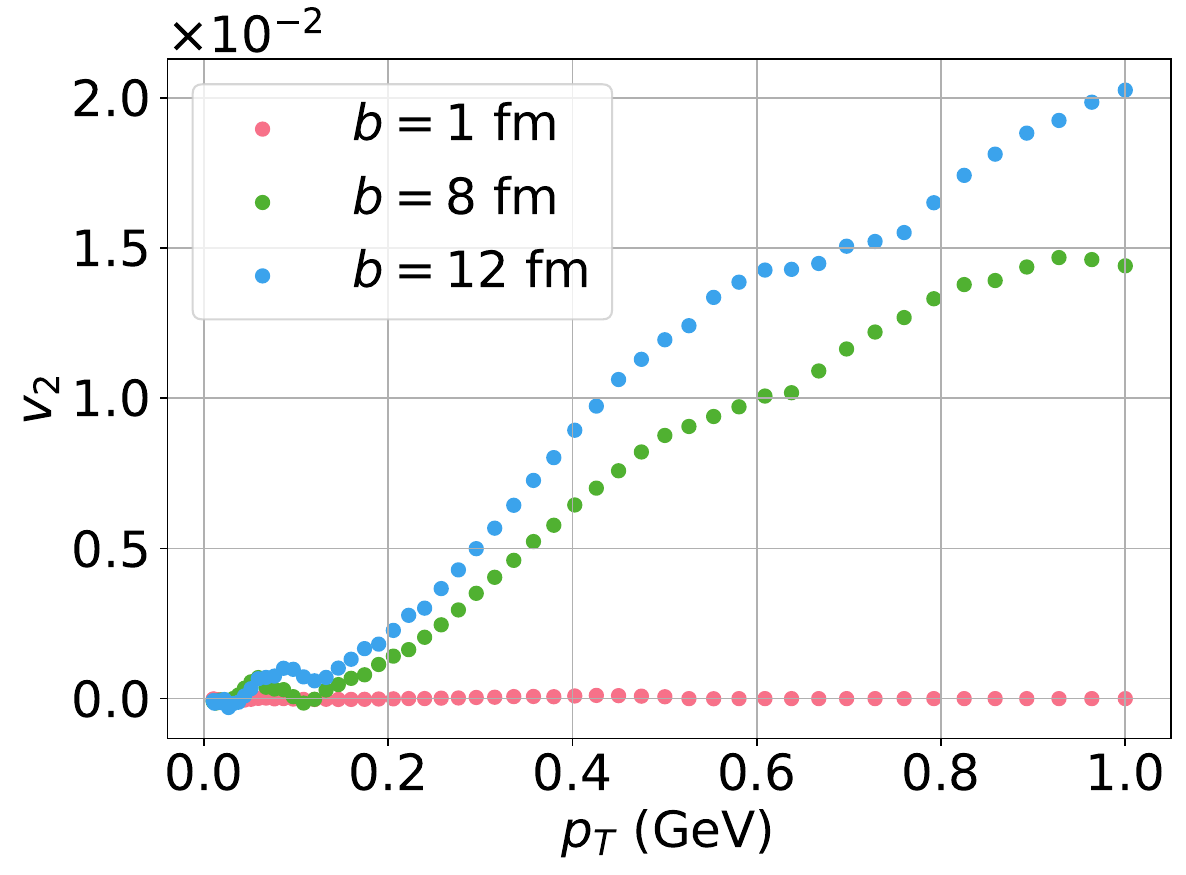}
        \label{fig:v2_b_anni}
    \end{subfigure}
    \hfill
    \begin{subfigure}{0.32\textwidth}
        \includegraphics[width=\textwidth]{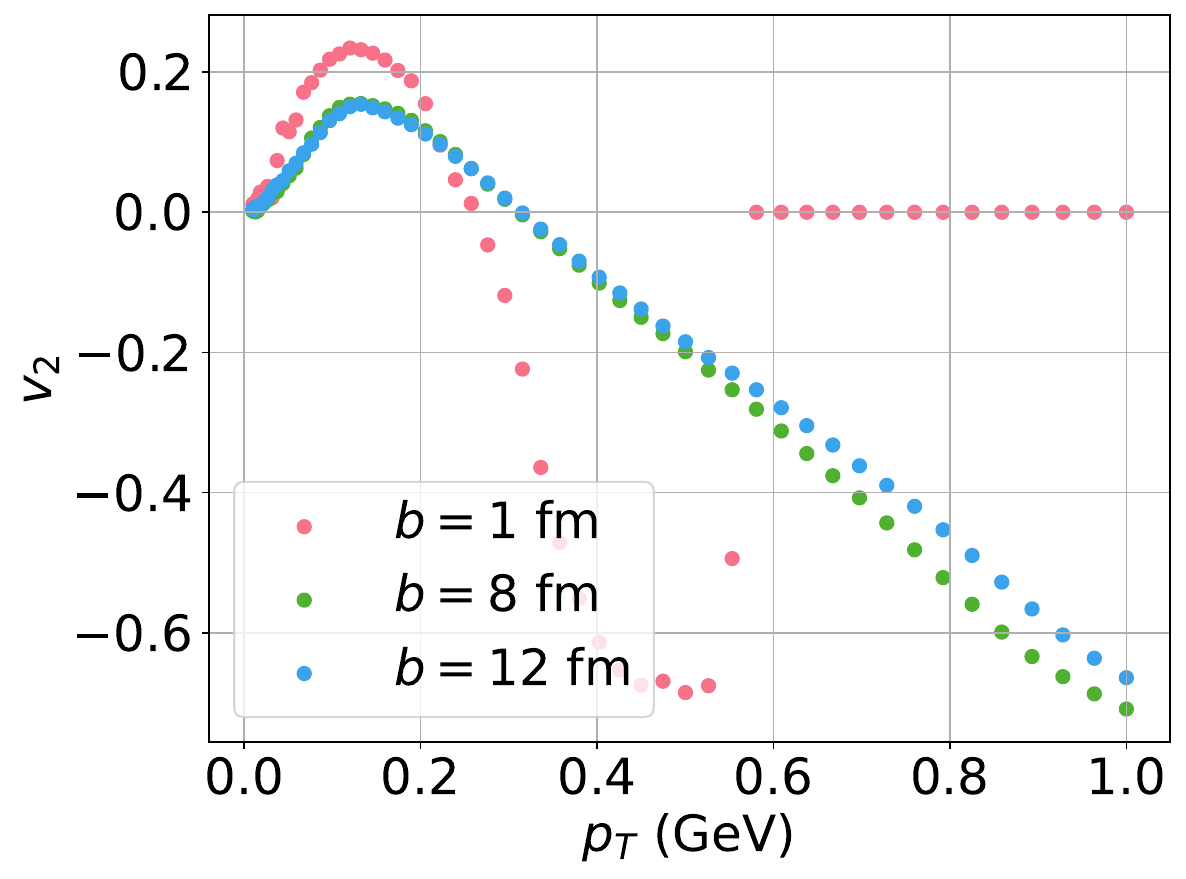}
        \label{fig:v2_b_decay}
    \end{subfigure}
    \caption{Elliptic flow coefficient $v_2$ as a function of transverse momentum $p_T$ for various impact parameters at given ($M, \sigma, Y$) = ($0.1$ GeV, $5.8$ MeV, 0). The left panel shows the total $v_2$, where as the other two panels show the effects of the annihilation (middle) and decay processes (right).}
    \label{fig:v2_comparison}
\end{figure}
In Fig.~\ref{fig:v2_comparison}, we present the dependence of elliptic flow ($v_2$) on the transverse momentum ($p_T$) for various impact parameters. From left to right, the panels show the total $v_2$, the contribution from annihilation, and the contribution from decay, respectively. Similar to the particle yield, the magnitude of the total $v_2$ increases with increasing magnetic field strength, i.e., from smaller to larger impact parameters. However, unlike the azimuthally integrated dilepton spectra, the decay processes individually exhibit a significantly larger elliptic flow compared to annihilation. At low $p_T$, the $v_2$ from decay processes decreases with increasing magnetic field, whereas for annihilation it always increases with magnetic field strength. 

\begin{figure}[hbtp]
    \centering
    \begin{subfigure}{0.45\textwidth}
        \includegraphics[width=\textwidth]{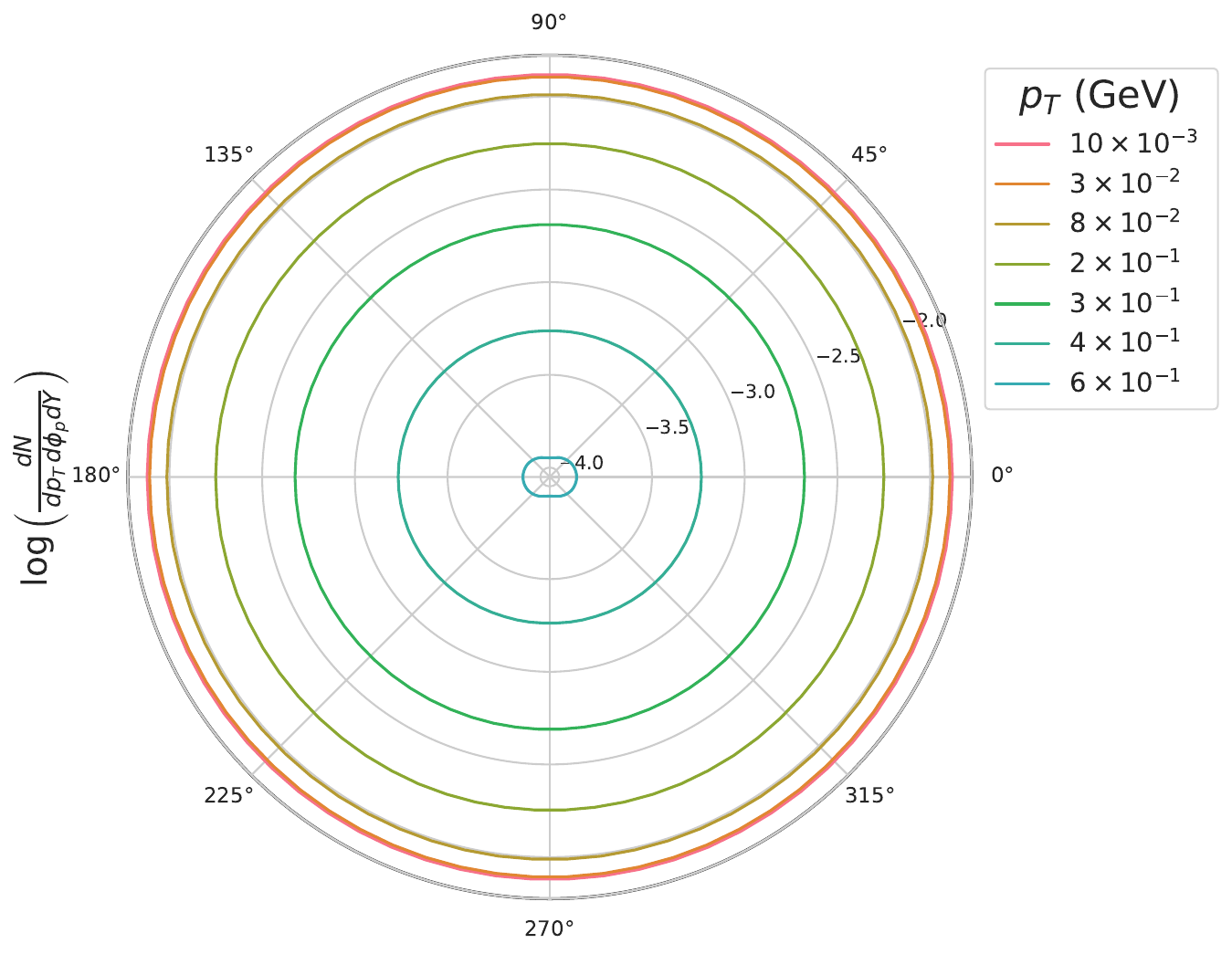}
         \label{fig:polar_anni}
    \end{subfigure}
    \hfill
    \begin{subfigure}{0.45\textwidth}
        \includegraphics[width=\textwidth]{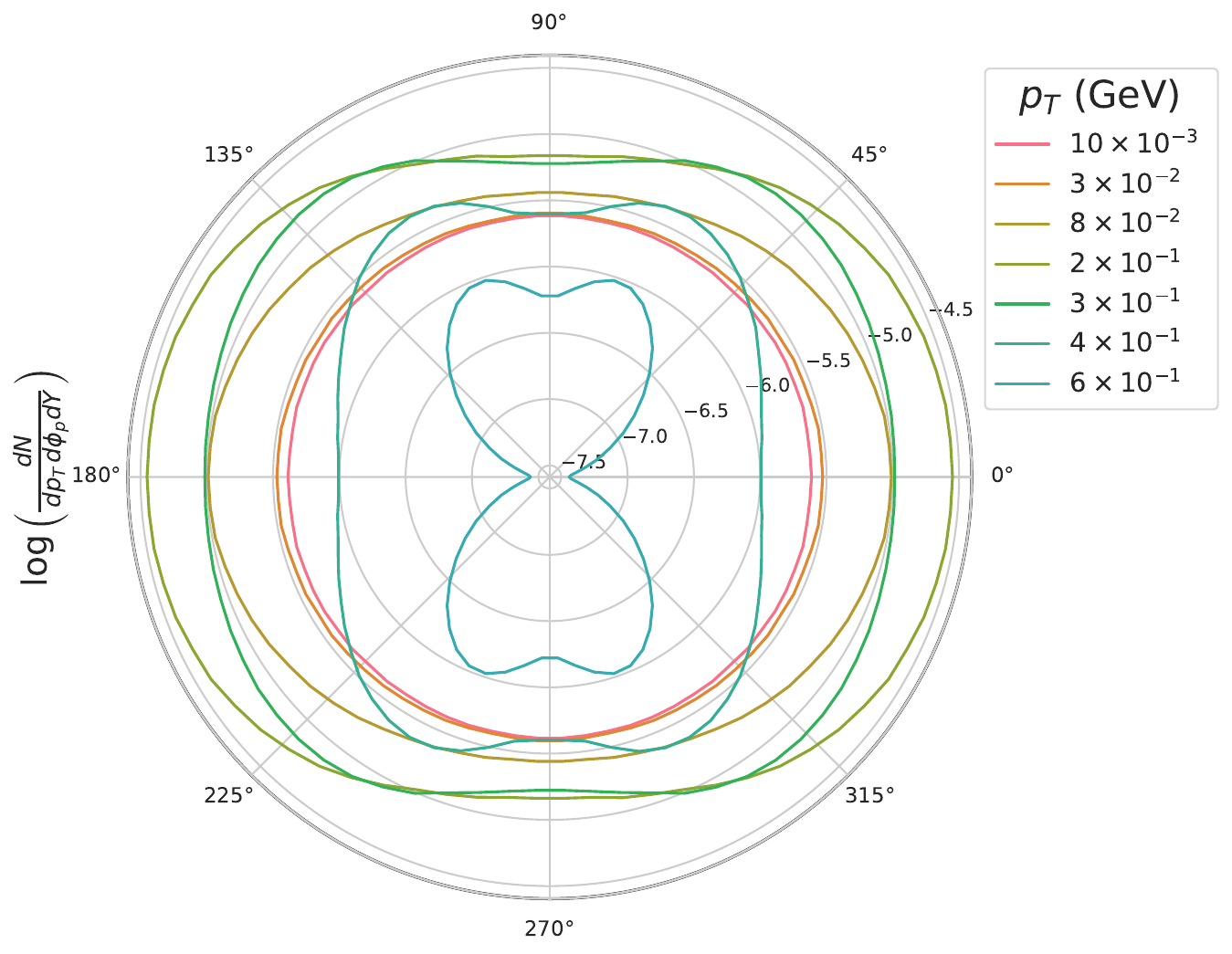}
        \label{fig:polar_decay}
    \end{subfigure}
    \caption{Polar plot of the azimuthal angle $\phi_p$  and transverse momentum $p_T$ dependence of the particle yield $\log\left({dN}/{(dp_T\,d\phi_p\,dY)}\right)$ at $(M, \sigma, b, Y) = (0.1~{\rm GeV}, 5.8~{\rm MeV}, 8~{\rm fm}, 0)$ for annihilation (left) and decay (right) processes respectively. The radial coordinate represents the $p_T$ values (indicated in the legend), while the angular coordinate corresponds to $\phi_p$.}
    \label{fig:polar_plots}
\end{figure}

Furthermore, the sign of $v_2$ for decay changes from positive at low $p_T$ to negative at high $p_T$, while for annihilation $v_2$ remains positive over the entire $p_T$ range. This behavior is illustrated in Fig.~\ref{fig:polar_plots}, which shows polar plots of the azimuthal angle $\phi_p$ and $p_T$ dependence of the particle yield for annihilation (left) and decay (right) processes. The transition from positive to negative $v_2$ for decay is reflected in the polar plots by a change in shape from  oblate (elongated along $x-$axis) to prolate (elongated along $y-$axis) as $p_T$ increases, whereas for the annihilation process, the shape starts from  an almost circular shape at low $p_T$ and becomes more oblate, as we increase $p_T$. 

An interesting observation from Fig.~\ref{fig:v2_comparison} is that even for $b = 1~\mathrm{fm}$, the $|v_2|$ from decay is the largest across all $p_T$ ranges (except at very high $p_T$, where the yield drops to zero). This is likely due to:  only a few cells have magnetic fields strong enough to contribute significantly, and hence the superposition of Landau level sums from different cells remains coherent (i.e., not spoiled) in this case. However, when computing the total $v_2$, the small yield from decay processes at low impact parameter, compared to annihilation, washes out the overall increase in $v_2$ from decay as evident from the left panel of Fig.~\ref{fig:v2_comparison}.

\begin{figure}[htbp]
    \centering
    \begin{subfigure}{0.32\textwidth}
        \includegraphics[width=\textwidth]{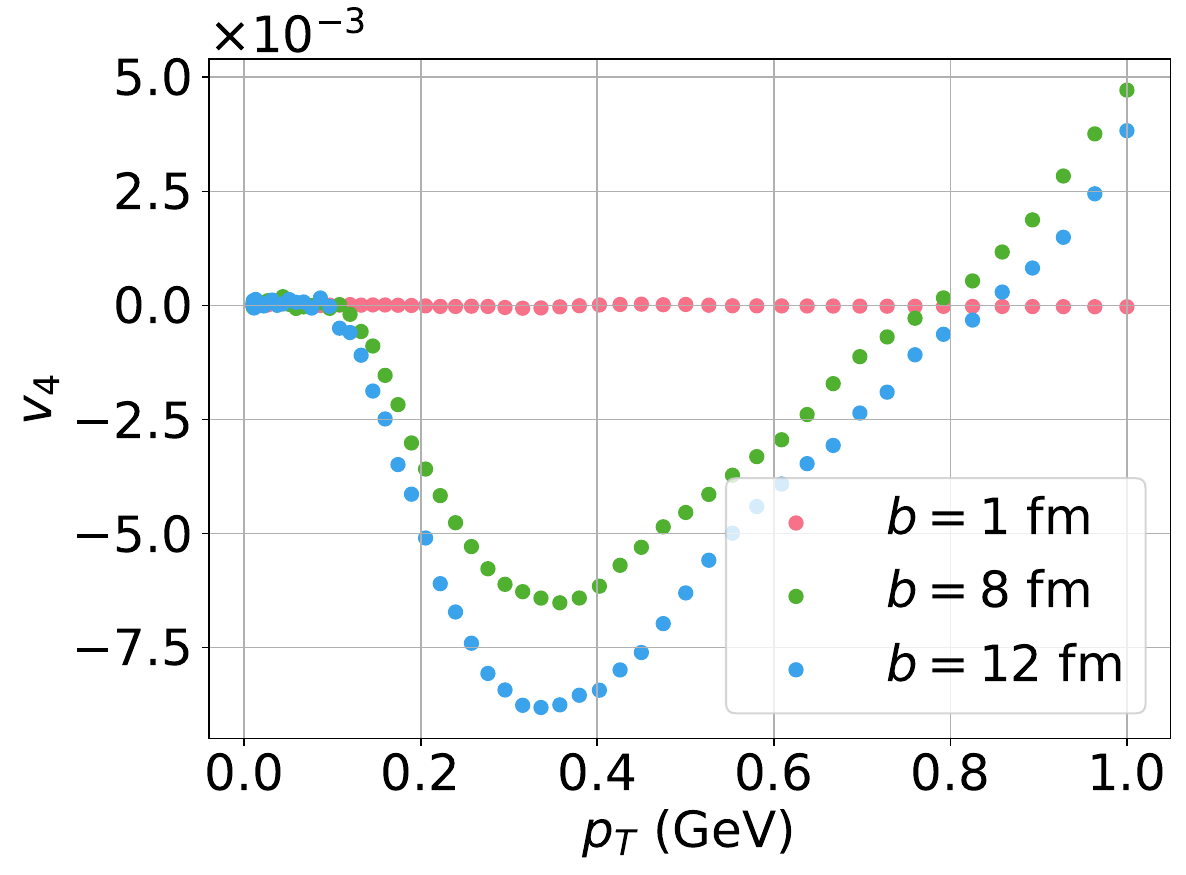}
        \label{fig:v4_b_total}
    \end{subfigure}
    \hfill
    \begin{subfigure}{0.32\textwidth}
        \includegraphics[width=\textwidth]{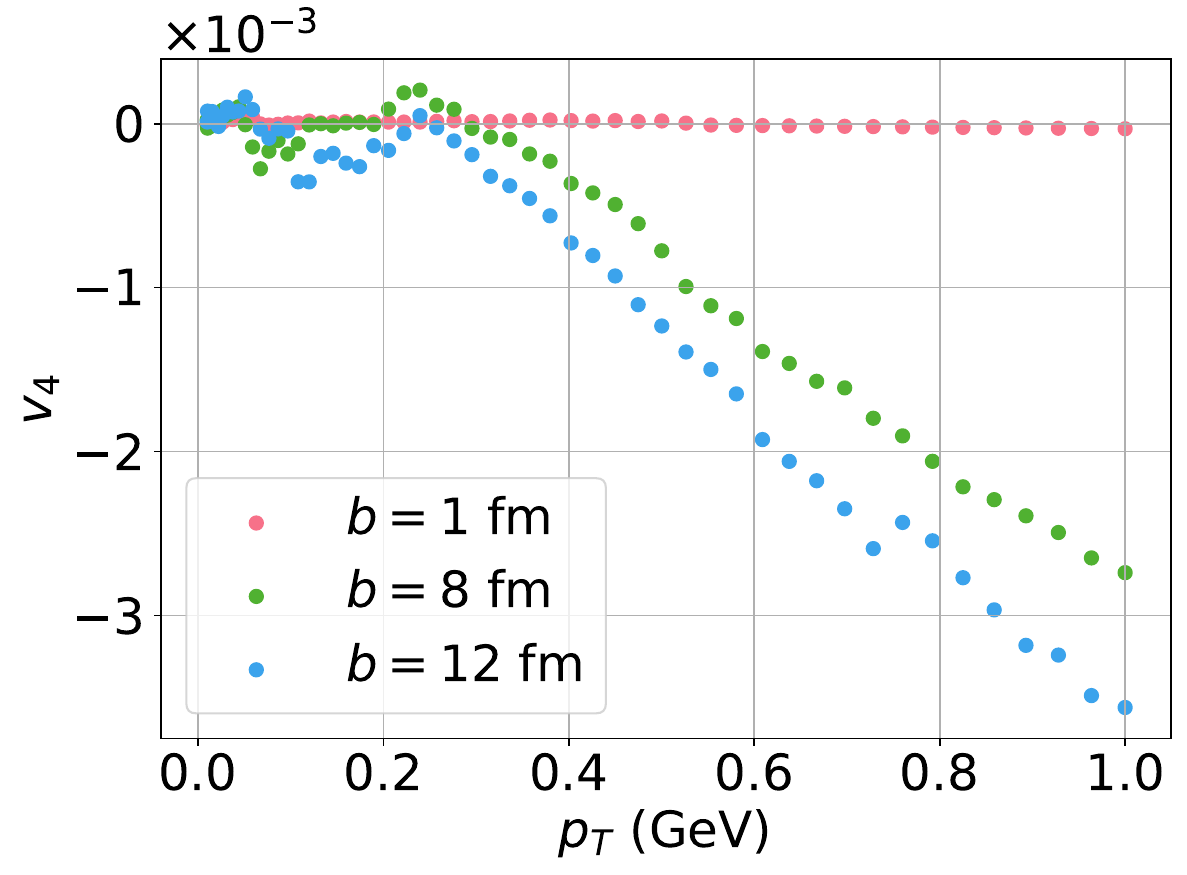}
        \label{fig:v4_b_anni}
    \end{subfigure}
    \hfill
    \begin{subfigure}{0.32\textwidth}
        \includegraphics[width=\textwidth]{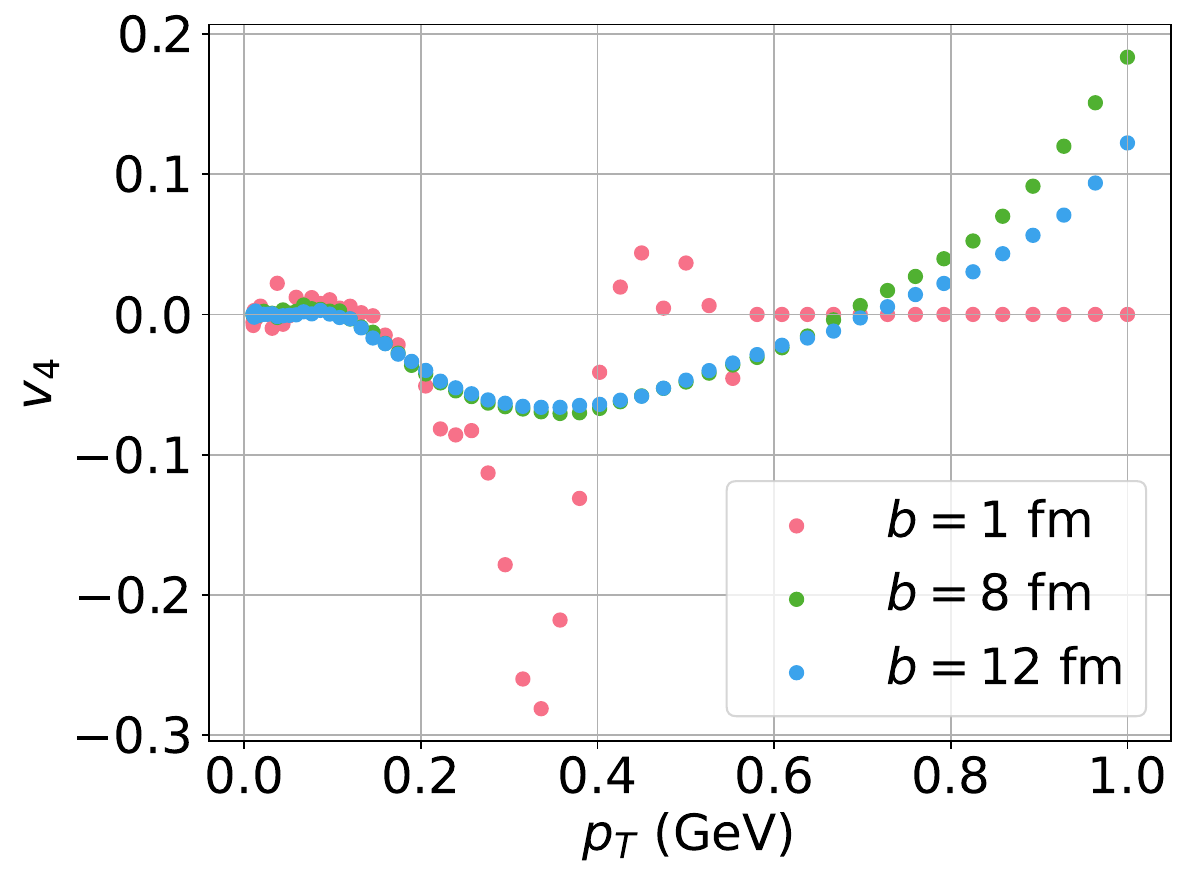}
        \label{fig:v4_b_decay}
    \end{subfigure}
    \caption{Quadrangular flow coefficient $v_4$ as a function of transverse momentum $p_T$ for various impact parameters at given ($M, \sigma, Y$) = ($0.1$ GeV, $5.8$ MeV, 0). The left panel shows the total $v_4$, where as the other two panels show the effects of the annihilation (middle) and decay (right) processes.}
    \label{fig:v4_comparison}
\end{figure}

A closer inspection of the polar plots in Fig.~\ref{fig:polar_plots} reveals that, in addition to the elliptic flow, there are clear traces of quadrangular flow $v_4$, particularly from the decay process. This coefficient, as a function of $p_T$, is shown in Fig.~\ref{fig:v4_comparison} for different impact parameters, in the same panel order as $v_2$. Similar to the elliptic flow, the total $v_4$ increases with impact parameter, although its magnitude is about an order smaller than $v_2$. 

Unlike the total $v_2$, which exhibits a single zero-crossing, the $v_4$ displays two zero-crossings—one at low $p_T$ and another at intermediate $p_T$. As anticipated, the $v_4$ from decay processes is about two orders of magnitude larger than that from annihilation processes. Moreover, at small impact parameter ($b = 1~\mathrm{fm}$), the $p_T$ dependence of $v_4$ has a characteristically different shape compared to that at large impact parameters. The origin of this difference can again be traced to the coherent superposition of Landau level sums. The effect of the cutoff parameter $\xi_{\mathrm{cf}}$ on the total $v_2(p_T)$ and $v_4(p_T)$ is discussed in Appendix~\ref{sec:cutoff}. 

\subsubsection{The effect of the lifetime of the magnetic field}
\label{sssec:eff_life_flow}
\begin{figure}[htbp]
    \centering
    \includegraphics[scale=0.4]
    {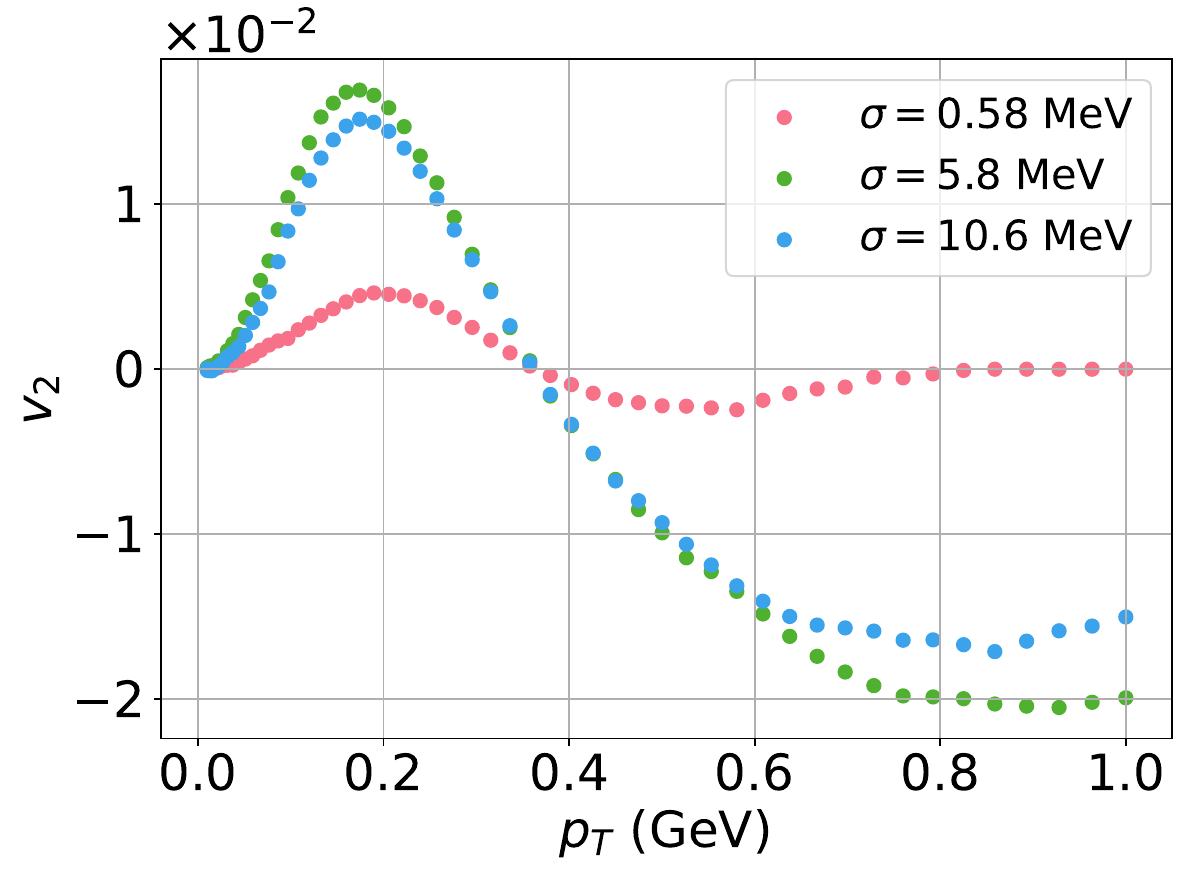}
    \includegraphics[scale=0.4]
    {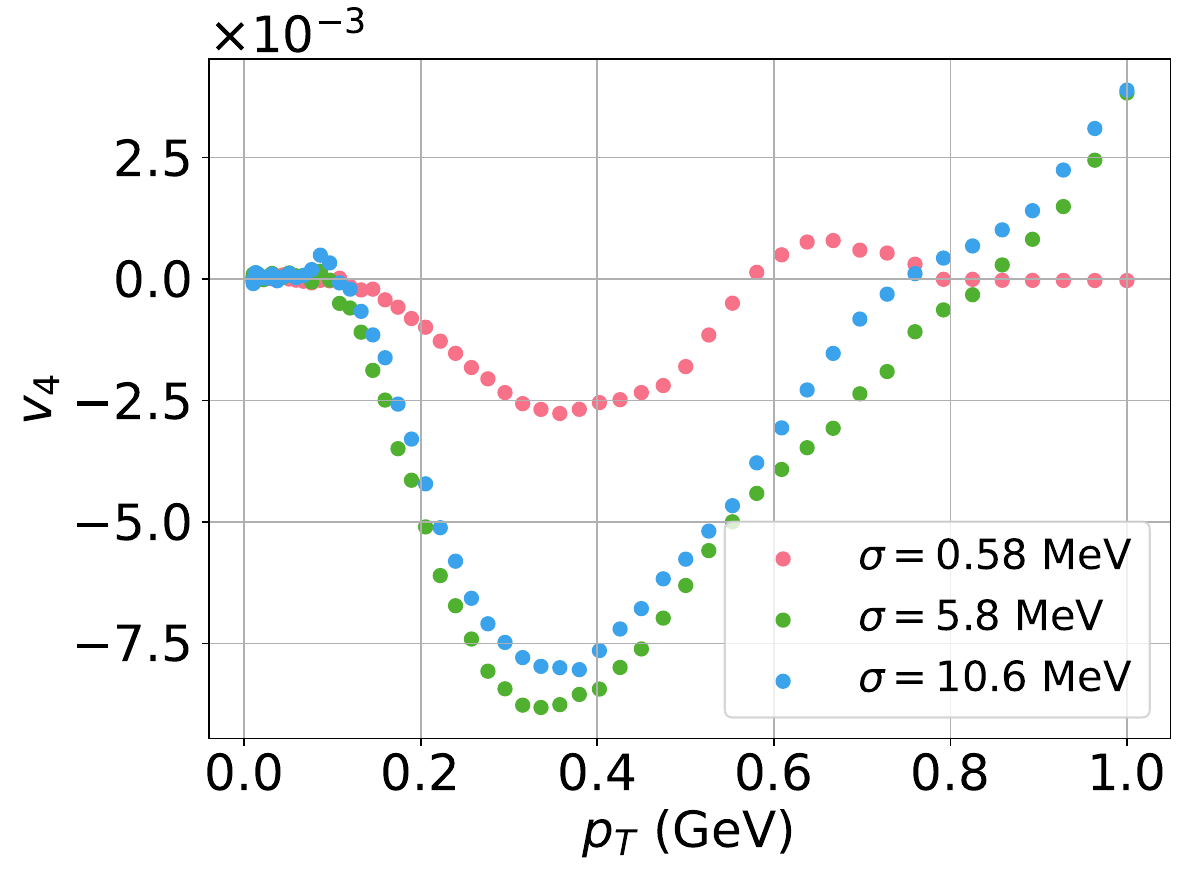}
    \caption{The total elliptic flow coefficient $v_2$ (left) and quadrangular flow coefficient $v_4$ (right) for various electrical conductivities at given $(M, b, Y)=(0.1~{\rm GeV}, 12~{\rm fm}, 0)$, illustrating the effect of the changing medium response.}
    \label{fig:v2v4_comparison_conductivities}
\end{figure}

Fig.~\ref{fig:v2v4_comparison_conductivities} illustrates how the magnetic field lifetime influences the even flow harmonics $v_2$ and $v_4$. The left panel shows the total elliptic flow $v_2$ as a function of $p_T$ for different conductivities. Comparing the vacuum case i.e. $\sigma =0.58$ MeV) to those with finite conductivity, we find that $v_2$ increases with conductivity, as expected, since a larger conductivity prolongs the lifetime of the magnetic field. However, when comparing the two finite-conductivity cases, a nontrivial trend emerges: the lower conductivity yields a larger $v_2$ than the higher one, similar to what was observed in Fig.~\ref{fig:spectra_comparison_conductivities}. As discussed earlier, this behavior should be regarded as a feature of the model rather than a genuine physical effect. Considering the processes separately (not shown here), we observed similar trends as for $v_2$ with respect to the impact parameter: the magnitude of $v_2$ from decay is larger than that from annihilation. Increasing the conductivity decreases $v_2$ for decay, while it increases $v_2$ for annihilation. However, when taking the weighted sum of the two contributions, the smaller yield from decay (refer  Fig.~\ref{fig:spectra_comparison} for comparison) causes the increase in $v_2$ from decay to be washed out. Other features, such as the zero-crossing at intermediate $p_T$, remain the same as discussed earlier.

The right panel of Fig.~\ref{fig:v2v4_comparison_conductivities} shows $v_4$ as a function of $p_T$ for various conductivities. The qualitative features are consistent with earlier observations: $v_4$ is about an order of magnitude smaller than $v_2$ and two zero-crossings are present across all conductivity values considered.

\subsubsection{The effect of the invariant mass}

\begin{figure}[htbp]
    \centering
    \includegraphics[scale=0.4]{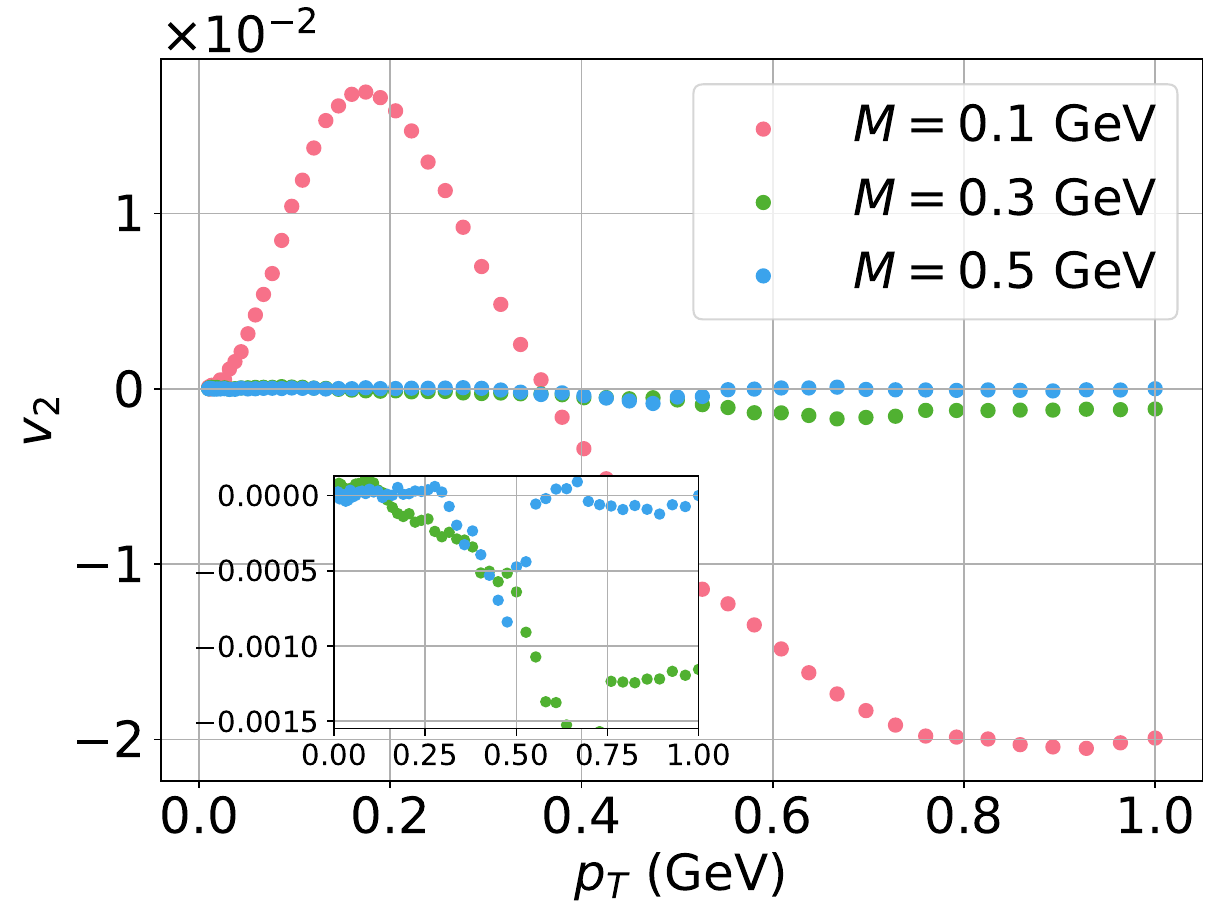}
    \includegraphics[scale=0.4]{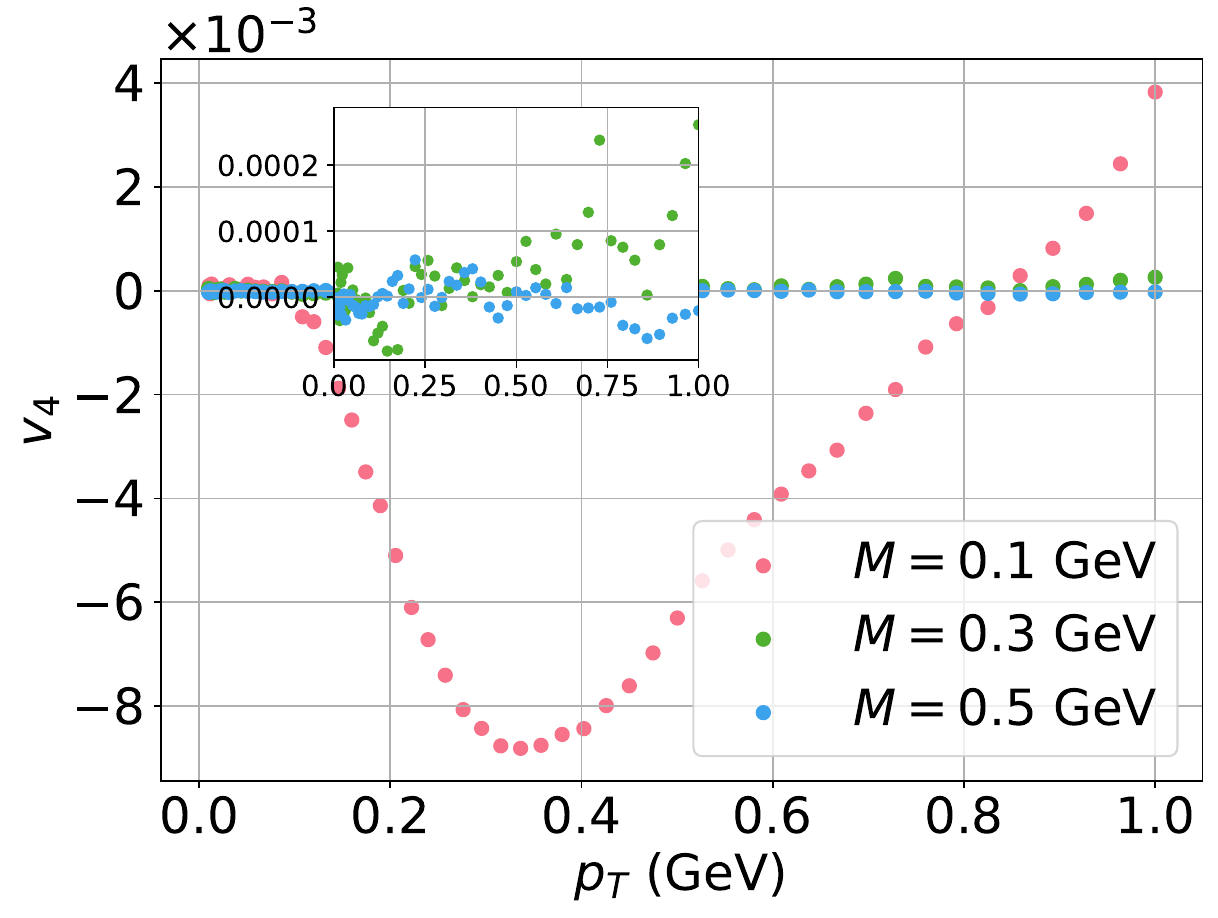}
    \caption{Total elliptic flow coefficient $v_2$ (left) and quadrangular flow coefficient $v_4$ (right)  for various invariant masses at fixed $(b,\sigma,Y)=(12~{\rm fm}, 5.8~{\rm MeV}, 0)$.}
    \label{fig:v2v4_comparison_masses}
\end{figure}

In the left panel of Fig.~\ref{fig:v2v4_comparison_masses} we have the total elliptic flow $v_2$ as a function of $p_T$ for different invariant masses. As the invariant mass increases, the magnitude of $v_2$ decreases across the entire $p_T$ range. In particular, the $v_2$ for $M=0.3~{\rm GeV}$ and $M=0.5~{\rm GeV}$ are about an order of magnitude smaller than those for $M=0.1~{\rm GeV}$. We also observe that for higher invariant masses, $v_2$ becomes negative even at very low transverse momenta ($p_T < 0.2~{\rm GeV}$), albeit with a small magnitude.  

The right panel of Fig.~\ref{fig:v2v4_comparison_masses} presents the corresponding quadrangular flow $v_4$ as a function of $p_T$. Similar to $v_2$, the magnitude of $v_4$ decreases with increasing invariant mass and is about an order smaller for $M=0.3~{\rm GeV}$ and $M=0.5~{\rm GeV}$ compared to $M=0.1~{\rm GeV}$. However, unlike $v_2$, the $v_4$ data points fluctuate strongly around zero over the full $p_T$ range, making it difficult to draw firm conclusions about its behavior.
 
\label{sssec:eff_mass_flow}

\section{Conclusion}
\label{sec:con}
The study in Ref.~\cite{Das:2021fma} demonstrated a pronounced enhancement of the dilepton production rate in the low invariant mass region. Moreover, the role of magnetic fields in generating anisotropies in photon and dilepton emission was explored in Refs.~\cite{Wang:2020dsr,Wang:2022jxx,Wang:2023fst}. However, these investigations were carried out in static backgrounds with spatially homogeneous, constant magnetic fields, often assuming unrealistically large field strengths. In realistic non-central HICs, by contrast, the magnetic field is typically much weaker---frequently below the characteristic pion-mass scale---and may thus yield negligible observable effects. This underscores the need for realistic space--time models of the QGP that account for expansion dynamics and spatially varying magnetic fields. Such modeling generally requires phenomenological frameworks like $3+1$D viscous hydrodynamic simulations~\cite{Schenke:2010rr, Karpenko:2013wva}, or ideally, $3+1$D magnetohydrodynamic (MHD) simulations~\cite{Nakamura:2022wqr,Mayer:2024dze,Mayer:2024kkv}.  

The present work aims to bridge this gap. Our approach is neither restricted to static backgrounds nor reliant on full $3+1$D MHD simulations. Instead, we incorporate realistic, inhomogeneous magnetic field profiles generated in HICs together with a QGP flow profile described by analytical Gubser’s solution to perform our calculations. A key advantage of using Gubser flow is that it enables us to disentangle effects arising from the collision geometry and medium flow from those intrinsic to the emission process itself.

In this study, we primarily varied two parameters: the impact parameter, which controls the strength of the magnetic field, and the conductivity, which affects the field's lifetime, at a fixed invariant mass. Subsequently, we fixed the impact parameter and conductivity and explored different invariant masses. Other variables, such as the center-of-mass energy and temperature etc., were held constant throughout. We found that the transverse momenta ($p_T$) spectra increase with increasing impact parameter, primarily due to the annihilation process, while the decay contribution remains sub-dominant. At small impact parameters, the combined (decay + annihilation) spectra match the Born rate. These observations, particularly the enhancement in the spectra with the magnetic field, are consistent with the expectations from the rate enhancement reported in Ref.~\cite{Das:2021fma}.

The most striking observations come from analyzing the elliptic flow across various impact parameters. We find a non-zero $v_2$ from decay processes even for nearly central collisions ($b=1$ fm). For the annihilation process, since the spectra are consistent with the Born rate, the harmonic contributions vanish. As the impact parameter increases, the elliptic flow from decay processes decreases, while that from annihilation increases. The magnitude of $|v_2|$ from decay is of order $\sim 0.1$, whereas that from annihilation is of order $\sim 10^{-2}$. When combined, the weighted $v_2$ is dominated by the annihilation contribution due to the larger particle yield, resulting in an overall smaller magnitude. Notably, $v_2$ from decay exhibits a characteristic shape: positive at small $p_T$ and negative at large $p_T$ across all impact parameters. In the polar plots of the yield as a function of $\phi_p$ and $p_T$, we found clear traces of quadrangular flow, $v_4$ from the decay process. Interestingly, $v_4$ from decay is comparable to $v_2$ in strength (though smaller than $v_2$), but with two zero crossings at large impact parameters compared to one for $v_2$.  

The spectra and harmonic flow as a function of conductivity show the expected behavior: larger conductivity leads to greater enhancement of the spectra and larger total elliptic flow. However, no distinctive features emerge that would allow for direct extraction of the conductivity from our analysis. For a fixed impact parameter and conductivity, varying the invariant mass reveals that the $p_T$ spectra are most enhanced for smaller invariant masses across all $p_T$ ranges, while for larger invariant masses, the effect is most pronounced at low $p_T$; at higher $p_T$, the spectra match the Born rate. Nevertheless, it is noteworthy that, contrary to naive expectations, even for the small magnetic fields considered in realistic simulations, this enhancement persists. For larger invariant masses ($M=0.3$ GeV and $M=0.5$ GeV), the magnitude of the harmonics is an order of magnitude smaller than for $M=0.1$ GeV. The elliptic flow for higher masses can become negative even at low $p_T$.  

Furthermore, our analysis indicates that central and semi-central collisions can play an important role in probing the electromagnetic fields produced in HICs. Although the  $v_2$ signal from decay processes is partially obscured by the background Born rate, separating the decay contribution would reveal a distinctive fingerprint of the background magnetic field in a hot QGP. Such a separation could also provide valuable insights into the dependence of $v_4$ on various parameters. Another key feature is the $p_T$ dependence of $v_2$, : it is positive at low $p_T$ and negative at high $p_T$ and hence is anti-correlated. This pattern is independent of impact parameter and conductivity and depends primarily on the invariant mass. Studying these correlations in more realistic scenarios, where the initial magnetic field and impact parameter vary event by event, will be crucial for future investigations aimed at probing the background magnetic field in HICs.
\\

{\bf Acknowledgments:} A.K.P. acknowledges the CSIR-HRDG financial support. Ar.D. acknowledges the support in part by the U.S. Department of Energy Grants No. DE-SC0023692. As.D. acknowledges support by the Deutsche Forschungsgemeinschaft (DFG, German Research Foundation) through the CRC-TR 211 “Strong-interaction matter under extreme conditions” – project number 315477589 – TRR 211. Computational resources have been provided by the Center for Scientific Computing (CSC) at the Goethe University. A.B. is supported by the European Union - NextGenerationEU through grant No. 760079/23.05.2023, funded by the Romanian ministry of research, innovation and digitalization through Romania’s National Recovery and Resilience Plan, call no. PNRR-III-C9-2022-I8. C.A.I. acknowledges the facilities provided by the University of Nova Gorica during his habilitation process.

{\bf Code Availability:} The source code and documentation can be found at the following GitHub repository:
\href{https://github.com/dashutosh573/Dileptons-Magneticfield}{\texttt{https://github.com/dashutosh573/Dileptons-Magneticfield}
}

\appendix
\section{Variable transformation and expression for Dilepton rate}
\label{sec:app_born}

In this section, we will show the transformation of the integration measure $d^4P$ from the usual momentum space $(p_0,{\mathbf p})$ to the one with variables such as the invariant mass $(M)$, transverse momentum $(p_T)$, azimuthal angle $(\phi_p)$ and rapidity $(Y)$. These are the variables in terms of which the results are expressed after performing the hydrodynamic evolution. Once the measure is expressed, we write down the expressions for the dilepton production rate for various processes. If we take the beam direction along the $z$-axis and the impact parameter along the $x$-axis, then 
\begin{align}
    M^2 = p_0^2 - \bm{p}^2,\quad p_x = p_T \cos\phi_p,\quad
    p_y = p_T \sin\phi_p, \quad  Y=\frac{1}{2}\ln\left(\frac{p_0+p_z}{p_0-p_z}\right).
\end{align}

Using the definition of rapidity, the four momentum variables can be written in terms of invariant mass, transverse momentum, rapidity and azimuthal angle by the following equation
\begin{align}
    p_0 = \sqrt{p_T^2+M^2}\cosh Y,
    \quad p_x = p_T \cos\phi_p, \quad p_y = p_T \sin\phi_p\quad {\rm and} \quad p_z = \sqrt{p_T^2+M^2}\sinh Y \label{eq:var_trans}
\end{align}
We need to know the Jacobian to write down the transformation from the momentum coordinates to the new coordinates,
\begin{align}
    d^4P=dp_0dp_xdp_ydp_z &= \left\vert\mathsf{det}\left[\mathbbm{J}\left(\frac{p_0,p_x,p_y,p_z}{M,p_T,Y,\phi_p}\right)\right]\right\vert dMdp_TdYd\phi_p
    =\left\vert\mathsf{det}\begin{pmatrix}
    \frac{\partial p_0}{\partial M} & \frac{\partial p_x}{\partial M} & \frac{\partial p_y}{\partial M} & \frac{\partial p_z}{\partial M} \\
    \frac{\partial p_0}{\partial p_T} & \frac{\partial p_x}{\partial p_T} & \frac{\partial p_y}{\partial p_T} & \frac{\partial p_z}{\partial p_T} \\
    \frac{\partial p_0}{\partial Y} & \frac{\partial p_x}{\partial Y} & \frac{\partial p_y}{\partial Y} & \frac{\partial p_z}{\partial Y} \\
    \frac{\partial p_0}{\partial \phi_p} & \frac{\partial p_x}{\partial \phi_p} & \frac{\partial p_y}{\partial \phi_p} & \frac{\partial p_z}{\partial \phi_p} 
    \end{pmatrix}\right\vert dMdp_TdYd\phi_p \nonumber \\
    & = M dM p_T dp_T dY d\phi_p
\end{align}

After finding out the transformations for the variables and the integration measure, we can now proceed to re-express the dilepton production rate for an arbitrarily magnetized medium evaluated in Ref.~\cite{Das:2021fma} in terms of the new coordinates. In this work, we assume the background magnetic field to be aligned along the $y$-axis, in contrast to Ref.~\cite{Das:2021fma}, where it was taken to point in the $z$-direction. So, here we define the parallel and perpendicular components\footnote{One must note the difference between the transverse $(p_T)$ and perpendicular components $(p_{\perp})$ of the momentum. The former lies on a plane perpendicular to the beam direction, while the latter lies on a plane perpendicular to the magnetic field direction.} of the momentum with respect to the $y$-axis,
\begin{align}
    p_{\parallel}^2 = p_0^2-p_y^2 = M_T^2\cosh^2 Y - p_T^2\sin^2\phi_p\quad {\rm and} \quad p_{\perp}^2 = p_x^2+p_z^2 = p_T^2\cos^2\phi_p + M_T^2\sinh^2 Y.
    \label{eq:pl_n_pp}
\end{align}
As written in Eq.~\eqref{eq:DPR_mag_total}, in the presence of an arbitrary background magnetic field, the dilepton production rate has three contributions\textemdash quark antiquark annihilation, quark decay and antiquark decay. We present their mathematical expressions in terms of the new variables, one by one. 

\subsection{Annihilation}
\begin{align}\label{eq:rate_ann}
\frac{dN}{d^4xMdMp_Tdp_TdYd\phi_p}\Bigg\vert_{q+\bar{q}\hookrightarrow \gamma^{*}} &= 4N_c\frac{\alpha^2_{\textsf{EM}}}{3\pi^3}\frac{1}{M^2}\frac{1}{\exp\left(\frac{M_T \cosh Y}{T}\right)-1}\sum_{f=u,d}\left(\frac{Q_f}{e}\right)^2\sum_{\ell,n = 0}^{\infty}(-1)^{\ell+n}\mathcal{N}_{\ell,n}\left(p^2_{\shp},p_{\sperp}^2\right)\\\nn
&\times\frac{\Theta\left(p_{\shp}^2-\left(m_{f,\ell}+m_{f,n}\right)^2\right)}{\lambda^{1/2}\left(p_{\shp}^2,m^2_{f,\ell},m^2_{f,n}\right)}\left[2-\tilde{n}_{+}\left(\mathcal{E}^{+}_{f,\ell,k}\right)
-\tilde{n}_{-}\left(\mathcal{E}^{+}_{f,n,q}\right)-\tilde{n}_{+}\left(\mathcal{E}^{-}_{f,\ell,k}\right)
-\tilde{n}_{-}\left(\mathcal{E}^{-}_{f,n,q}\right)\right],
\end{align}
where, $m_{f,\ell}=\sqrt{2 \ell |Q_fB|+m_f^2}$ is the Landau level dependent mass.
There are restrictions on the number of Landau levels, both on the indices $n$ and $l$ in Eq.~\ref{eq:rate_ann} and are given as, $n\leq \ell +\left\lfloor\frac{p_{\shp}^2-2m_{f,\ell}\sqrt{p_{\shp}^2}}{2|Q_{\sF}B|}\right\rfloor$ and $\ell \leq \left\lfloor\frac{p_{\shp}^2-2m_{\sF}\sqrt{p_{\shp}^2}}{2|Q_{\sF}B|}\right\rfloor$, respectively. The brackets are used to indicate the floors.

\subsection{Particle Decay}
\begin{align}
\label{eq:rate_decay_p}
\frac{dN}{d^4xMdMp_Tdp_TdYd\phi_p}\Bigg\vert_{q\rightarrow q+\gamma^*} &=4N_c\frac{\alpha^2_{\textsf{EM}}}{3\pi^3}\frac{1}{M^2}\frac{1}{\exp\left(\frac{M_T \cosh Y}{T}\right)-1}\sum_{f=u,d}\left(\frac{Q_f}{e}\right)^2\sideset{}{'}\sum_{\ell,n=0}^{\infty}(-1)^{\ell+n}\mathcal{N}_{f,\ell,n}\left(p^2_{\shp},p_{\sperp}^2\right)\\\nn
&\times\frac{\Theta\left(\left(m_{f,\ell}-m_{f,n}\right)^2-p_{\shp}^2\right)}{\lambda^{1/2}\left(p_{\shp}^2,m^2_{f,\ell},m^2_{f,n}\right)}\left[\tilde{n}_{+}\left(\mathcal{E}^{+}_{f,\ell,k}\right)-\tilde{n}_{+}\left(-\mathcal{E}^{+}_{f,n,q}\right)+\tilde{n}_{+}\left(\mathcal{E}^{-}_{f,\ell,k}\right)-\tilde{n}_{+}\left(-\mathcal{E}^{-}_{f,n,q}\right)\right].
\end{align} 
The prime in the summation over the Landau level denotes the condition $\ell >n$. In this case,  the condition $\left(m_{f,\ell}-m_{f,n}\right)^2-p_{\shp}^2 \geq 0$ and $\ell > n$ lead to the upper limit of of $n$ as $n\leq \ell +\left\lfloor\frac{p_{\shp}^2-2m_{f,\ell}\sqrt{p_{\shp}^2}}{2|Q_{\sF}B|}\right\rfloor$ and the lower limit of $\ell$ as $\ell \geq \left\lfloor\frac{p_{\shp}^2+2m_{\sF}\sqrt{p_{\shp}^2}}{2|Q_{\sF}B|}\right\rfloor$.

\subsection{Anti-particle Decay}
\begin{align}
\label{eq:rate_decay_ap}
\frac{dN}{d^4xMdMp_Tdp_TdYd\phi_p}\Bigg\vert_{\bar{q}\rightarrow \bar{q}+\gamma^*} &=4N_c\frac{\alpha^2_{\textsf{EM}}}{3\pi^3}\frac{1}{M^2}\frac{1}{\exp\left(\frac{M_T \cosh Y}{T}\right)-1}\sum_{f=u,d}\left(\frac{Q_f}{e}\right)^2\sideset{}{''}\sum_{\ell,n=0}^{\infty}(-1)^{\ell+n}\mathcal{N}_{f,\ell,n}\left(p^2_{\shp},p_{\sperp}^2\right)\\\nn
&\times\frac{\Theta\left(\left(m_{f,\ell}-m_{f,n}\right)^2-p_{\shp}^2\right)}{\lambda^{1/2}\left(p_{\shp}^2,m^2_{f,\ell},m^2_{f,n}\right)}\left[\tilde{n}_{-}\left(\mathcal{E}^{+}_{f,n,q}\right)-\tilde{n}_{-}\left(-\mathcal{E}^{+}_{f,\ell,k}\right)+\tilde{n}_{-}\left(\mathcal{E}^{-}_{f,n,q}\right)-\tilde{n}_{-}\left(-\mathcal{E}^{-}_{f,\ell,k}\right)\right].
\end{align}
Note that here in the summation over the Landau level, the condition $n>\ell$ must be satisfied and is denoted by a double prime. The upper limit of $\ell$ is $\ell\leq n + \left\lfloor\frac{p_{\shp}^2-2m_{f,n}\sqrt{p_{\shp}^2}}{2|Q_{\sF}B|}\right\rfloor$. On the other hand, the lower limit of $n$ is $n \geq \left\lfloor\frac{p_{\shp}^2+2m_{\sF}\sqrt{p_{\shp}^2}}{2|Q_{\sF}B|}\right\rfloor.$\\

Finally, to execute those three equations we need to know a few terms. We list them below.
\begin{itemize}
\item We have already given $\{p_{\shp},p_{\sperp}\}$ in Eq.~\ref{eq:pl_n_pp} in terms of $\{M_T,p_T,Y,\phi_p\}$ and can be easily converted into $\{M,p_T,Y,\phi_p\}$ using $M_T=\sqrt{M^2+p^2_T}$. 

\item The $\theta$-function, $\Theta\left(p_{\shp}^2-\left(m_{f,\ell}+m_{f,n}\right)^2\right)$ provides the condition on the momentum, and needs to be expressed as $\Theta\left(M_T^2\cosh^2 Y - p_T^2\sin^2\phi_p-\left(m_{f,\ell}+m_{f,n}\right)^2\right)$. 

\item We need to express the other terms to express the whole expression in terms of the desired variables, namely, $\mathcal{N}_{\ell,n}\left(p^2_{\shp},p_{\sperp}^2\right)$, 
$\lambda^{1/2}\left(p_{\shp}^2,m^2_{f,\ell},m^2_{f,n}\right)$ and the distribution functions $\tilde{n}_{\pm}\left(\mathcal{E}^{\pm}_{\alpha,\beta,\gamma}\right)$. They are given as,
\begin{align}
\mathcal{N}_{f,\ell,n}(p_{\shp}^2,p_{\sperp}^2) &\equiv \left[(\ell+n)|Q_{\sF}B|-\frac{1}{2}p_{\shp}^2\right]\left[\mathcal{J}^{(0)}_{f,\ell-1,n}(p_{\sperp}^2) + \mathcal{J}^{(0)}_{f,\ell,n-1}(p_{\sperp}^2)\right]+m_{\sF}^2\left[\mathcal{J}^{(0)}_{f,\ell,n}(p_{\sperp}^2)
+\mathcal{J}^{(0)}_{f,\ell-1,n-1}(p_{\sperp}^2)\right]\nn\\
&+8\mathcal{J}_{f,\ell-1,n-1}^{(1)}(p_{\sperp}^2),
\label{eq:def_I}
\end{align}
where 
\begin{align}
\mathcal{J}^{(\alpha)}_{f,\ell,n}(p_{\sperp}^2) \equiv \int\!\!\frac{d^2k_{\sperp}}{(2\pi)^2}\exp\left(-\frac{k_{\sperp}^2+p_{\sperp}^2}{|Q_{\sF}B|}\right)\,(k.p)_{\sperp}^{\alpha}\,L_{\ell}^{\alpha}\left(\frac{2k_{\sperp}^2}{|Q_{\sF}B|}\right)L_{n}^{\alpha}\left(\frac{2k_{\sperp}^2}{|Q_{\sF}B|}\right). 
\label{eq:Jalpha}
\end{align}
At first appearance, the expression might appear involved, but it is not that complicated as we look deep into it. The integration given in the above Eq.~\ref{eq:Jalpha} can be performed analytically. This has been performed and given in detail in the appendix of Ref.~\cite{Das:2021fma}.

To evaluate Eq.~\ref{eq:def_I}, we need to know Eq.~\ref{eq:Jalpha} for $\alpha =0\,\, {\rm and}\,\, 1$. They are given below.
\begin{align}
\mathcal{J}^{(0)}_{f,\ell,n}(p_{\sperp}^2) = 
\begin{cases}
(-1)^{\ell+n}\dfrac{|Q_{\sF}B|}{8\pi}\dfrac{\textsf{min}(\ell,n)!}{\textsf{max}(\ell,n)!}\xi^{|\ell-n|}e^{-\xi}\left[L_{\textsf{min}(\ell,n)}^{|\ell-n|}(\xi)\right]^2 & \text{for  } \,\,\,p_{\sperp}\neq 0\\
\dfrac{|Q_{\sF}B|}{8\pi}\delta_{\ell, n} & \text{for  }\,\,\, p_{\sperp} = 0  
\end{cases} \label{eq:J0}
\end{align}
and
\begin{align}
\mathcal{J}^{(1)}_{f,\ell,n}(p_{\sperp}^2) =
\begin{cases}
 (-1)^{\ell+n}\dfrac{|Q_{\sF}B|^2}{16\pi}\dfrac{\big(\textsf{min}(\ell,n)+1\big)!}{\textsf{max}(\ell,n)!}\xi^{|\ell-n|}e^{-\xi}L_{\textsf{min}(\ell,n)}^{|\ell-n|}(\xi)L_{\textsf{min}(\ell,n)+1}^{|\ell-n|}(\xi) & \text{for  } \,\,\,p_{\sperp}\neq 0\\
\dfrac{|Q_{\sF}B|^2}{16\pi}(\ell+1)\delta_{\ell+1, n+1} & \text{for  } \,\,\,p_{\sperp} = 0\,, 
 \end{cases} \label{eq:J1}
\end{align}
for  $\alpha =0\, {\rm and}\, 1$, respectively; $\xi = \dfrac{p_{\sperp}^2}{2|Q_{\sF}B|}$. $L_{\ell}^{\alpha} (x)$ is the generalized Laguerre polynomial given as
\begin{align}
(1-z)^{-(\alpha+1)}\exp\left(\frac{xz}{z-1}\right) = \sum_{\ell =0}^{\infty} L_{\ell}^\alpha(x) z^{\ell}. \label{eq:Laguerre_generating}
\end{align}

\item Next, we express the other term, $\lambda^{1/2}\left(p_{\shp}^2,m^2_{f,\ell},m^2_{f,n}\right)$. This is known as the K{\" a}ll{\' e}n function (or to be precise square root of that) and is given as
\begin{align}
\lambda^{1/2}(p_{\shp}^2,m^2_{f,\ell},m^2_{f,n})=\left(p_{\shp}^4+m^4_{f,\ell}+m^4_{f,n}-2p_{\shp}^2m^2_{f,\ell}-2m^2_{f,\ell}m^2_{f,n}-2m^2_{f,n}p_{\shp}^2\right)^{1/2}.
\end{align}

\item Now, only the distribution functions are left to be expressed. $\wt{n}_{\pm}(\mathcal{E})$ is the Fermi-Dirac distribution
\begin{align}
\wt{n}_{\pm}(\mathcal{E}) = \frac{1}{\exp(\frac{\mathcal{E}\mp\mu}{T})+1}
\end{align}

The arguments $\mathcal{E}^{\pm}_{f,\ell,k}$ and $\mathcal{E}^{\pm}_{f,n,q}$ are given by
\begin{align}
\mathcal{E}^{\sigma}_{f,\ell,k} = s_1\frac{p_0\left(p_{\shp}^2+2 (\ell-n)|Q_fB|\right)+\sigma |p_z|\lambda^{1/2}(p_{\shp}^2,m_{f,\ell}^2,m_{f,n}^2)}{2p_{\shp}^2}.
\end{align}
and
\begin{align}
\mathcal{E}^{\sigma}_{f,n,q} = s_2\frac{p_0\left(p_{\shp}^2+2 (n-\ell)|Q_fB|\right)-\sigma |p_z|\lambda^{1/2}(p_{\shp}^2,m_{f,\ell}^2,m_{f,n}^2)}{2p_{\shp}^2},
\end{align} 
with $\sigma=\pm1$. 

\end{itemize}

For the sake of completeness, we also provide the expression of the leading order dilepton rate in absence of any magnetic field, the so called Born rate~\cite{McLerran:1984ay},
\begin{align}
    &\frac{dN}{d^4xMdMp_Tdp_TdYd\phi_p}\Bigg\vert_{\rm Born} \nonumber \\
    &=\frac{5N_c\alpha^2_{\textsf{EM}}}{108\pi^4}\frac{1}{\sqrt{M^2_T\sinh^2Y+p^2_T}}\frac{T}{e^{\frac{M_T\cosh Y}{T}}-1}\left(1+\frac{2m_f^2}{M^2}\right)\log\left(\frac{\left\{\exp\left(-\frac{M_T\cosh Y}{T}\right)+e^{-\omega_-/T}\right\}e^{-\omega_+/T}}{\left\{\exp\left(-\frac{M_T\cosh Y}{T}\right)+e^{-\omega_+/T}\right\}e^{-\omega_-/T}}\right),
\label{eq:born}
\end{align}
where $\omega_{\pm} = \frac{M_T}{2}\left(\cosh Y \pm \sqrt{\sinh^2 Y+\frac{p_T^2}{M_T^2}}\sqrt{1-\frac{4m_f^2}{M^2}}\right)$.

\section{Numerical convergence and sensitivity of spectra and harmonics to cutoff artifacts}
\label{sec:cutoff}
\begin{figure}[htbp]
    \centering
    \begin{subfigure}{0.32\textwidth}
        \includegraphics[width=\textwidth]{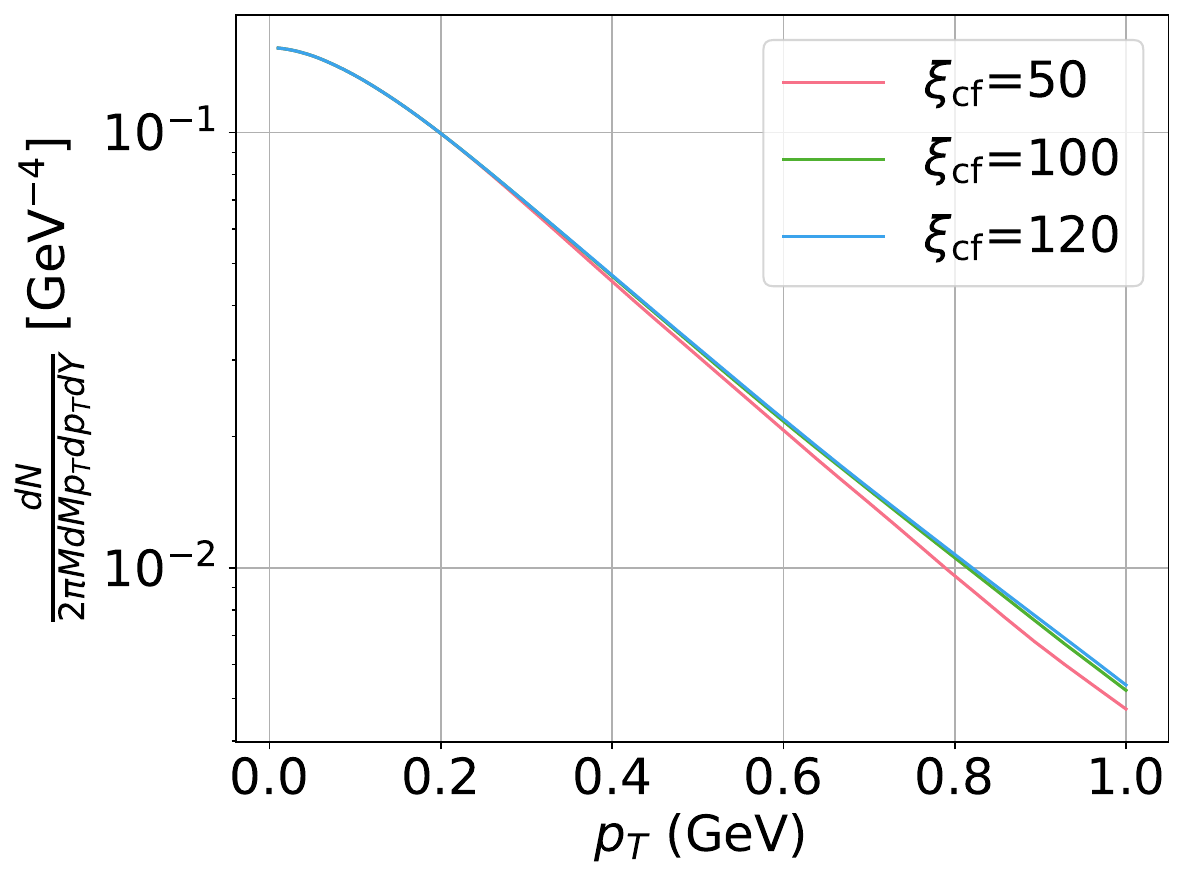}
    \end{subfigure}
    \hfill
    \begin{subfigure}{0.32\textwidth}
        \includegraphics[width=\textwidth]{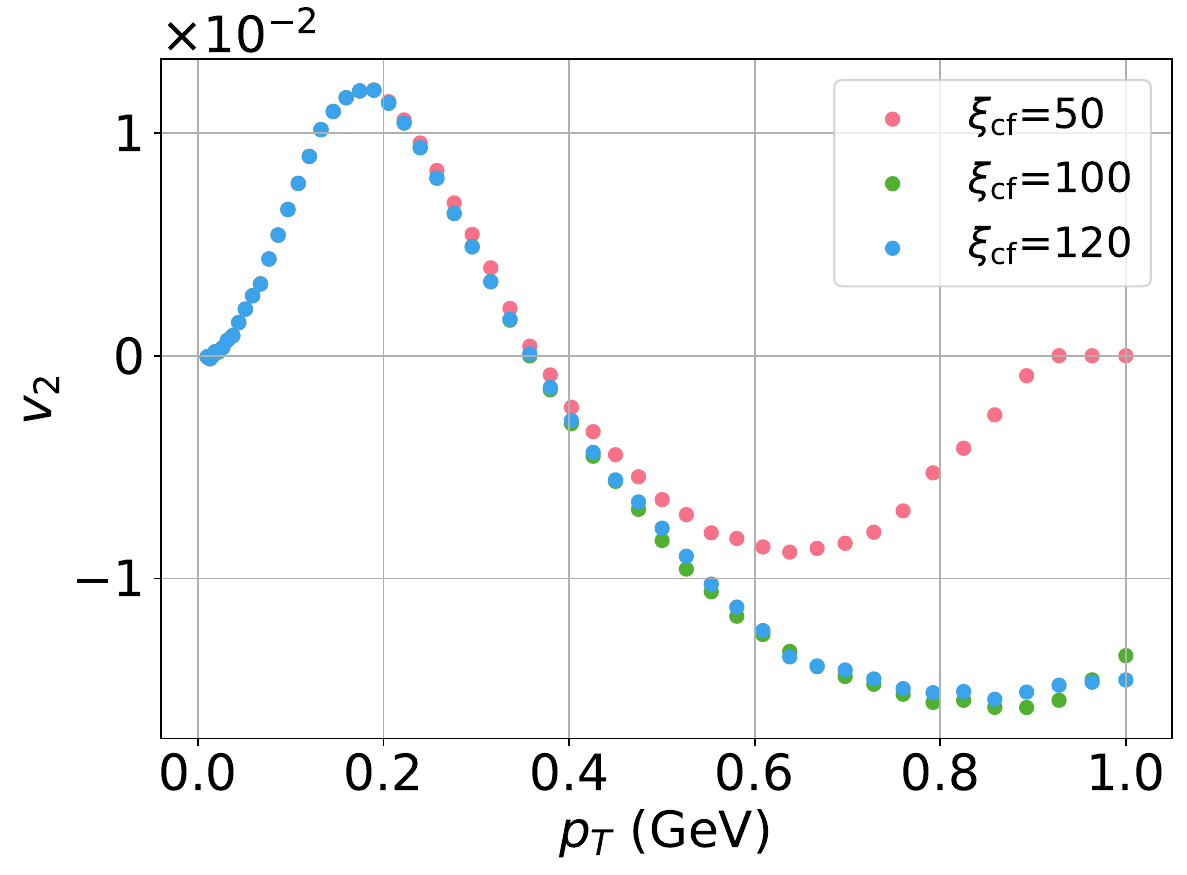}
    \end{subfigure}
    \hfill
    \begin{subfigure}{0.32\textwidth}
        \includegraphics[width=\textwidth]{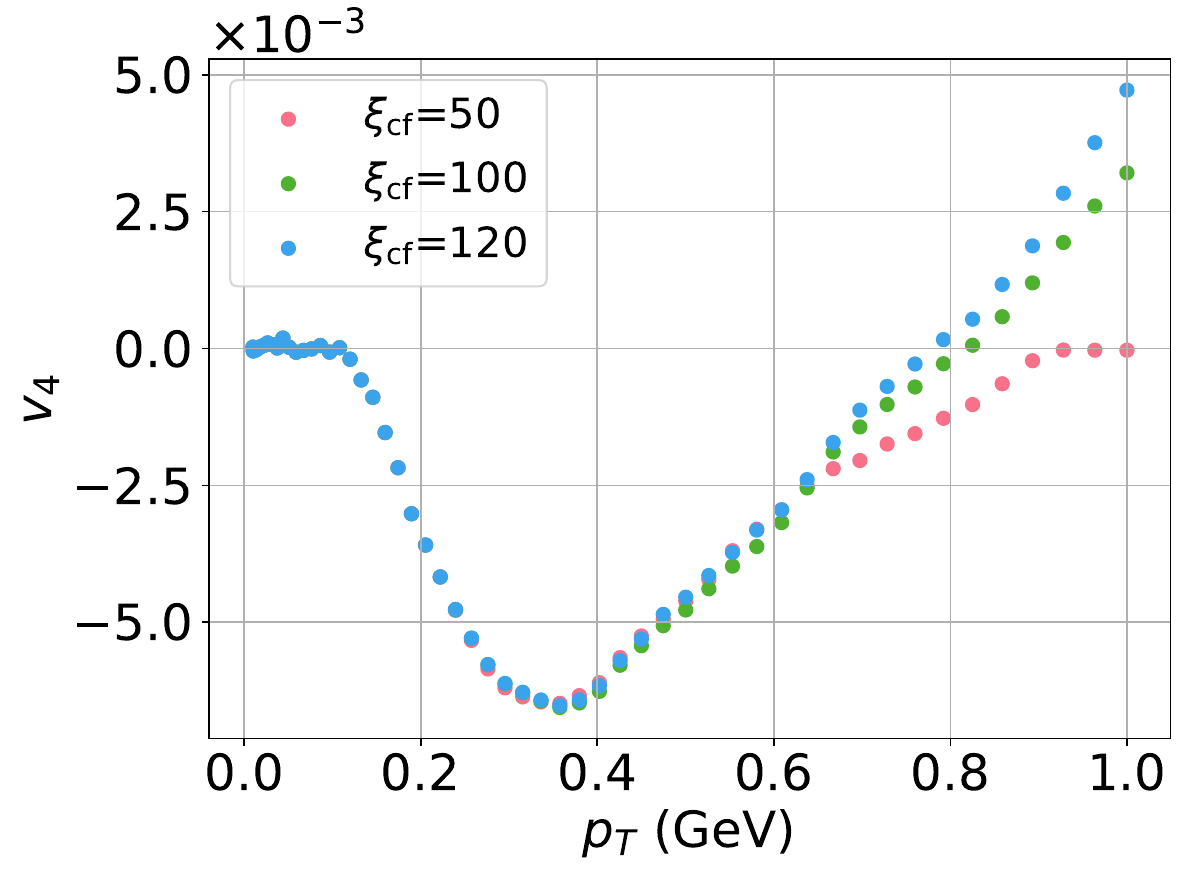}
    \end{subfigure}
    \caption{ Numerical sensitivity of total spectra (left), $v_2$ (middle) and $v_4$ (right) for $b = 8$ fm, $\sigma = 5.8$ MeV and $M = 100$ MeV for three different choices of cutoffs $\xi_{\mathrm{cf}}$= 50, 100, 120 respectively.}
    \label{fig:sensitivty_comparision}
\end{figure}
As discussed in Subsection~\ref{subsec:numerical_setup}, we introduce a cutoff parameter $\xi_{\mathrm{cf}}$ to stabilize the numerics for small magnetic fields, since the Laguerre polynomials grow rapidly at large $p_\perp$ for such small  values of magnetic fields. It is therefore important to assess the impact of this cutoff on observable quantities.  

In Fig.~\ref{fig:sensitivty_comparision}, we compare the sensitivity of the spectra (left), $v_2(p_T)$ (middle), and $v_4(p_T)$ (right) for different choices of the cutoff parameter $\xi_{\mathrm{cf}}$. As shown, the spectra, $v_2(p_T)$, and $v_4(p_T)$ exhibit excellent agreement in the low transverse momentum region $p_T \lesssim 0.6$~GeV. In particular, for $v_2$, small deviations appear even after $p_T \gtrsim 0.4$~GeV.  Nevertheless, it should be stressed that such small cutoffs $\xi_{\mathrm{cf}}=50$ has only been used in case of larger invariant mass. Therefore, all numerical data points beyond these ranges should be interpreted with caution, particularly for large invariant masses.

\bibliography{ref_common}

\end{document}